\documentclass[]{aastex63}

\graphicspath{{./}{figures/}}
\newcommand{\tess}{{\it TESS}}

\newcommand{\kepler}{{\it Kepler}}
\newcommand{\gaia}{{\it Gaia}}
\newcommand{\ktwo}{{K2}}
\newcommand{\spitzer}{{Spitzer}}
\newcommand{\vizier}{VizieR}
\newcommand{\programname}{\texttt{PROGRAMNAME}}
\newcommand{\degs}{$^{\circ}$}
\newcommand{\teff}{$T_{\rm eff}$}
\newcommand{\logg}{$\log{g}$}

\newcommand{\beatonetal}{R.~Beaton et al. (submitted; AAS29028)}
\newcommand{\beatonetalpar}{ R.~Beaton et al. submitted; AAS29028}
\newcommand{\kms}{$\,{\rm km}\,{\rm s}^{-1}\:$}

\shorttitle{APOGEE-2S Targeting}
\shortauthors{Santana et al.}

\begin{document}
\title{Final Targeting Strategy for the SDSS-IV APOGEE-2S Survey}
\correspondingauthor{Felipe A. Santana}
\email{fsantana@das.uchile.cl}
\author[0000-0002-4023-7649]{Felipe~A.~Santana} \affiliation{Departamento de
Astronom\'ia, Universidad de Chile, Camino El Observatorio 1515, Las Condes, Santiago}
%
%
\author[0000-0002-1691-8217]{Rachael~L.~Beaton}
\altaffiliation{Much of this work was completed while this author was a NASA Hubble Fellow at Princeton University.}
\altaffiliation{Carnegie-Princeton Fellow}
\affiliation{Department of Astrophysical Sciences, Princeton University, 4 Ivy Lane, Princeton, NJ~08544}
\affiliation{The Observatories of the Carnegie Institution for Science, 813 Santa Barbara St., Pasadena, CA~91101}
\author[0000-0001-6914-7797]{Kevin~R.~Covey} 
\affiliation{Department of Physics \& Astronomy, Western Washington University, Bellingham, WA, 98225, USA}
%
%
\author[0000-0003-2321-950X]{Julia~E.~O\rq{}Connell}
\affiliation{Departamento de Astronom{\'{\i}}a, Universidad de Concepci{\'o}n, Casilla 160-C, Concepci{\'o}n, Chile}
%
%
\author{Pen\'elope~Longa-Pe\~na} 
\affiliation{Unidad de Astronom\'ia, Universidad de Antofagasta, Avenida Angamos 601, Antofagasta 1270300, Chile}
%
%
\author{Roger Cohen} 
\affiliation{Space Telescope Science Institute, Baltimore, MD, 21218, USA}
\author{Jos\'e~G.~Fern\'andez-Trincado} 
\affiliation{Instituto de Astronom\'ia y Ciencias Planetarias, Universidad de Atacama, Copayapu 485, Copiap\'o, Chile}
\author[0000-0003-2969-2445]{Christian~R.~Hayes} 
\affiliation{Department of Astronomy, University of Washington, Seattle, WA, 98195, USA}
\author{Gail~Zasowski}\affiliation{Department of Physics \& Astronomy, University of Utah, Salt Lake City, UT, 84112, USA}
%
%
\author{Jennifer~S.~Sobeck} \affiliation{Department of Astronomy, University of Washington, Seattle, WA, 98195, USA}
\author{Steven~R.~Majewski} 
\affiliation{Department of Astronomy, University of Virginia, Charlottesville, VA, 22903, USA}
%
%
\author{S.~D.~Chojnowski} 
\affiliation{Department of Astronomy, New Mexico State University, Las Cruces, NM, 88001, USA}
\affiliation{Department of Physics, Montana State University, P.O. Box 173840, Bozeman, MT 59717-3840, USA}
\author{Nathan De Lee} \affiliation{Department of Physics, Geology, and Engineering Technology, Northern Kentucky University, Highland Heights, KY 41099, USA}
\affiliation{Department of Physics \& Astronomy, Vanderbilt University, Nashville, TN, 37235, USA}
%
\author{Ryan~J.~Oelkers}
\affiliation{Department of Physics \& Astronomy, Vanderbilt University, Nashville, TN, 37235, USA}
\author[0000-0003-1479-3059]{Guy~S.~Stringfellow} \affiliation{Center for Astrophysics and Space Astronomy, Department of Astrophysical and Planetary Sciences, University of Colorado, Boulder, CO, 80309, USA}
%
%
\author{Andr\'es~Almeida}
\affiliation{Instituto de Investigaci\'on Multidisciplinario en Ciencia y Tecnolog\'ia, Universidad de La Serena. Avenida Ra\'ul Bitr\'an S/N, La Serena, Chile}
\author{Borja~Anguiano} \affiliation{Department of Astronomy, University of Virginia, Charlottesville, VA, 22903, USA}
\affiliation{Department of Physics \& Astronomy, Macquarie University, Balaclava Rd, NSW 2109, Australia}
\author{John~Donor} \affiliation{Department of Physics \& Astronomy, Texas Christian University, Fort Worth, TX, 76129, USA}
\author[0000-0002-0740-8346]{Peter~M.~Frinchaboy} \affiliation{Department of Physics \& Astronomy, Texas Christian University, Fort Worth, TX, 76129, USA}
\author{Sten~Hasselquist}\affiliation{Department of Physics \& Astronomy, University of Utah, Salt Lake City, UT, 84112, USA} 
\affiliation{NSF Astronomy and Astrophysics Postdoctoral Fellow} 
\author{Jennifer~A.~Johnson} \affiliation{Department of Astronomy, The Ohio State University, Columbus, OH, 43210, USA}
\author{Juna~A.~Kollmeier}\affiliation{The Observatories of the Carnegie Institution for Science, 813 Santa Barbara St., Pasadena, CA~91101}
\author{David~L.~Nidever}
\affiliation{Department of Physics, Montana State University, P.O. Box 173840, Bozeman, MT 59717-3840, USA}
\author[0000-0003-0872-7098]{Adrian.~M.~Price-Whelan}
\affiliation{Center for Computational Astrophysics, Flatiron Institute, 162 Fifth Ave, New York, NY 10010, USA}
\author{Alvaro Rojas-Arriagada}
\affiliation{Instituto de Astrof\'{i}sica, Facultad de F\'{i}sica, Pontificia Universidad Cat\'{o}lica de Chile, Av. Vicu\~{n}a Mackenna 4860, Santiago, Chile}
\affiliation{Millennium Institute of Astrophysics, Av. Vicu\~{n}a Mackenna 4860, 782-0436, Macul, Santiago, Chile}
\author{Mathias Schultheis}
\affiliation{Observatoire de la Cote d'Azur, Lagrange Boulevard de l'Observatoire 06304 Nice, France}
\author{Matthew~Shetrone}
\affiliation{University of California Observatories, Santa Cruz, CA 95064, USA}
\author{Joshua~D.~Simon}\affiliation{The Observatories of the Carnegie Institution for Science, 813 Santa Barbara St., Pasadena, CA~91101}
%
%
\author{Conny Aerts} \affiliation{Institute of Astronomy, KU Leuven, Celestijnenlaan 200D, B-3001 Leuven, Belgium}
\author[0000-0002-0786-7307]{Jura~Borissova}
\affiliation{Instituto de F\'isica y Astronom\'ia, Universidad de Valpara\'iso, Av. Gran Breta\~na 1111, Playa Ancha, Casilla 5030, Chile.}
\affiliation{Millennium Institute of Astrophysics, Av. Vicu\~{n}a Mackenna 4860, 782-0436, Macul, Santiago, Chile}
\author{Maria~R.~Drout}
\affiliation{The Observatories of the Carnegie Institution for Science, 813 Santa Barbara St., Pasadena, CA~91101}
\affiliation{Department of Astronomy and Astrophysics, University of Toronto, 50 St. George Street, Toronto, Ontario, M5S 3H4 Canada}
\author{Doug~Geisler}
\affiliation{Departamento de Astronom{\'{\i}}a, Universidad de Concepci{\'o}n, Casilla 160-C, Concepci{\'o}n, Chile}
\affiliation{Instituto de Investigaci\'on Multidisciplinario en Ciencia y Tecnolog\'ia, Universidad de La Serena. Avenida Ra\'ul Bitr\'an S/N, La Serena, Chile}
\affiliation{Departamento de Astronomia, Facultad de Ciencias, Universidad de La Serena. Av. Juan Cisternas 1200, La Serena, Chile}
\author{C.Y. Law}   
\affiliation{Department of Space, Earth \& Environment, Chalmers University of Technology, SE-412 96 Gothenburg, Sweden}
\affiliation{Department of Physics, The Chinese University of Hong Kong, Shatin, NT, Hong Kong SAR}
\author{Nicolas~Medina}
\affiliation{Instituto de F\'isica y Astronom\'ia, Universidad de Valpara\'iso, Av. Gran Breta\~na 1111, Playa Ancha, Casilla 5030, Chile.}
\affiliation{Millennium Institute of Astrophysics (MAS), Santiago, Chile.}
\author[0000-0002-7064-099X]{Dante~Minniti}
\affiliation{Departamento de Ciencias F\'isicas, Facultad de Ciencias Exactas, Universidad Andr\'es Bello, Fern\'andez Concha 700, Las Condes, Santiago, Chile}
\affiliation{Vatican Observatory, Vatican City State, V-00120, Italy}
\author[0000-0003-2325-9616]{Antonela~Monachesi}
\affiliation{Instituto de Investigaci\'on Multidisciplinario en Ciencia y Tecnolog\'ia, Universidad de La Serena. Avenida Ra\'ul Bitr\'an S/N, La Serena, Chile}
\affiliation{Departamento de Astronomia, Facultad de Ciencias, Universidad de La Serena. Av. Juan Cisternas 1200, La Serena, Chile}
\author{Ricardo\ R.\ Mu\~noz} \affiliation{Departamento de Astronom\'ia, Universidad de Chile, Camino El Observatorio 1515, Las Condes, Santiago}
\author[0000-0002-9245-6368]{Rados\l{}aw Poleski}
\affiliation{Astronomical Observatory, University of Warsaw, Al. Ujazdowskie 4, 00-478 Warszawa, Poland}
\affiliation{Department of Astronomy, The Ohio State University, Columbus, OH, 43210, USA}
\author[0000-0002-1379-4204]{Alexandre~Roman-Lopes}
\affiliation{Departamento de Astronomia, Facultad de Ciencias, Universidad de La Serena. Av. Juan Cisternas 1200, La Serena, Chile}
\author[0000-0001-5761-6779]{Kevin C. Schlaufman}
\affiliation{Department of Physics and Astronomy, Johns Hopkins
University, 3400 N Charles St., Baltimore, MD 21218, USA}
\author{Amelia~M.~Stutz}
\affiliation{Departamento de Astronom{\'{\i}}a, Universidad de Concepci{\'o}n, Casilla 160-C, Concepci{\'o}n, Chile}
\author{Johanna~Teske}\altaffiliation{NASA Hubble Fellow} 
\affiliation{Earth and Planets Laboratory, Carnegie Institution for Science, 5241 Broad Branch Road, NW, Washington, DC 20015, USA}
\altaffiliation{Much of this work was completed while this author was a NASA Hubble Fellow at the Observatories of the Carnegie Institution for Science.}
\author{Andrew~Tkachenko} \affiliation{Institute of Astronomy, KU Leuven, Celestijnenlaan 200D, B-3001 Leuven, Belgium}
\author{Jennifer~L.~Van Saders}\affiliation{Institute for Astronomy, University of Hawai\rq{}i, 2680 Woodlawn Drive, Honolulu, HI 96822, USA}
\author{Alycia~Weinberger}
\affiliation{Earth and Planets Laboratory, Carnegie Institution for Science, 5241 Broad Branch Road, NW, Washington, DC 20015, USA}
\author{Manuela~Zoccali}
\affiliation{Millennium Institute of Astrophysics, Av. Vicu\~{n}a Mackenna 4860, 782-0436, Macul, Santiago, Chile}
\affiliation{Instituto de Astrof\'isica, Pontificia Universidad Cat\'olica de Chile, Av. Vicu\~na Mackenna 4860, 782-0436 Macul, Santiago, Chile}

%

%


%
\begin{abstract}
APOGEE is a high-resolution ($R\sim22,000$), near-infrared, 
multi-epoch, spectroscopic survey of the Milky Way.
The second generation of the APOGEE project, APOGEE-2, includes 
an expansion of the survey to the Southern Hemisphere called 
APOGEE-2S. This expansion enabled APOGEE to perform a fully
panoramic mapping of all the main regions of the Milky Way; in particular, by operating in the $H$-band, APOGEE is uniquely able to probe the  dust-hidden inner regions of the Milky Way that are best accessed from the Southern Hemisphere.
In this paper we present the targeting strategy of APOGEE-2S,
with special attention to documenting modifications to the original, previously published plan.
The motivation for these changes is explained as well as an 
assessment of their effectiveness in achieving their intended scientific objective. 
In anticipation of this being the last paper detailing APOGEE targeting, we present an accounting of
all such information complete through the end of the APOGEE-2S
project; this includes several main survey programs dedicated
to exploration of major stellar populations and regions of the Milky Way, as well as a full list of
programs contributing to the APOGEE database through allocations of observing time by the Chilean National Time Allocation Committee (CNTAC) and the Carnegie
Institution for Science (CIS).
\added{This work was presented along with a companion article, \beatonetal, presenting the final target selection strategy adopted for APOGEE-2 in the Northern Hemisphere.}
\end{abstract}


\section{Introduction}
\label{sec:intro}
\defcitealias{zasowski2017}{Z17}
\defcitealias{zasowski2013}{Z13}

Detailed analyses of the different substructures of the Milky Way
and the physical processes that govern their interplay require
high precision, homogeneous, large data samples covering all of
its main regions. In that sense, modern 
photometric surveys like the Sloan Digital Sky Survey 
\citep[SDSS;][]{york2000} or the Two Micron All-Sky Survey 
\citep[2MASS;][]{skrutskie2006} have been key to expanding our 
knowledge about our Milky Way. 
The remarkable discoveries enabled by these SDSS and 2MASS imaging include: 
    (i) a large population of ultrafaint dwarf galaxies and low luminosity star clusters  \citep{belokurov2007,koposov2007,belokurov2010},
    (ii) constraints on the morphology of both the stellar and dark matter halo \citep[e.g.,][]{majewski2003,helmi2004,johnston2005,law2005,belokurov2006,juric2008,lm10,deg2013}, 
    (iii) tidal tails from both disrupting star clusters and Milky Way satellite galaxies \citep[e.g.,][]{ivezic2000,odenkirchen2001,newberg2002,majewski2003,grillmair2006a,grillmair2006b,grillmair2009}, 
    (iv) multiple stellar populations in globular clusters \citep[e.g.,][]{lardo2011,gratton2012}, and 
    (v) the characterization of complex substructures in the bulge/bar association using red clump stars \citep[e.g.,][]{mcwilliam2010,saito2011,wegg2015}.

Despite the importance of photometric surveys in addressing these topics, full 
chemo-dynamical models of the Milky Way require input from precise and
large-scale spectroscopic datasets.
In this sense, a large number of recent spectroscopic Galactic 
surveys have enabled mapping the large phase space covered by 
the different stellar populations of the Milky Way,
and the intrinsic properties of those stars, e.g., their masses, temperatures, chemistry and evolutionary
stages.
Some examples of the contributions from these surveys have been the new estimates of escape velocity and mass of the Milky Way by \citet{piffl2014} using data from the Radial Velocity Experiment \citep[RAVE;][]{steinmetz2006}, and the constraints on
the formation and evolution of the thin disk, thick disk and halo \citep{lee2011,schlaufman2009,schlaufman2011,schlaufman2012} using data from the Sloan Extension for Galactic Understanding and Exploration survey \citep[SEGUE;][]{yanny2009}.

Most major spectroscopic surveys of the Milky Way performed to date, including 
    SEGUE \citep[][C.~Rokosi et al.~(in prep.)]{yanny2009},
    RAVE \citep[][]{steinmetz2006},
    \gaia-ESO \citep{gilmore2012}
    LAMOST \citep{cui2012}, and
    GALAH \citep{galah_overview}
    use the optical range of the electromagnetic spectrum. 
As a result, their sampling of both the inner Galaxy and within the Galactic plane is
extremely limited due to the severe attenuation of optical light by dust.
However, the Galactic mid-plane, inner disk and Milky Way bulge are critical to understanding Galactic evolution because these regions of highest stellar density contain the vast majority of the Milky Way's stellar mass as well as stellar populations spanning the largest range in chemical composition and age.

The need for a panoramic, homogeneous, high-precision, spectroscopic study throughout all the major components of the Milky Way motivated the inception of the Apache Point Galactic Evolution Experiment \citep[APOGEE;][]{majewski2017}.
APOGEE\added{\footnote{The nomenclature we will use is as follows: APOGEE refers to the joint APOGEE-1 and APOGEE-2 project or tools common to it, APOGEE-N or APOGEE-S refer to the spectrographs in the Northern and Southern hemisphere, respectively, and APOGEE-1 or APOGEE-2 refer to the SDSS-III and SDSS-IV surveys, respectively.}} is a high-resolution ($R\sim22,500$), near-infrared (NIR; $\lambda=1.51$--$1.70\,{\rm \mu m}$), multi-epoch, spectroscopic survey of the Milky Way.
During its first generation (2011-2014; hereafter APOGEE-1), the survey was one of the projects within SDSS-III \citep{eisenstein2011}, and provided spectra for $\sim163,000$ stars with a signal-to-noise ratio ($S/N$) of $\sim100$, covering a large variety of Galactic environments.
Along with the spectra, APOGEE-1 provided high level data products derived from the spectra --- fundamental stellar parameters like radial velocity, effective temperature, surface gravity, and chemical abundances for 15 elements, the latter derived via the APOGEE Stellar Parameter and Chemical Abundance Pipeline \citep[ASPCAP;][]{garciaperez2016}; \deleted{which} \added{the cumulative results from APOGEE-1} were released in DR12 \citep{holtzman2015}.

APOGEE data have been applied to a wide variety of topics in stellar astrophysics and Galactic archaeology.  
APOGEE-enabled research covers a wide range of Galactic environments allowing for the detailed study of the Milky Way's main structures: 
    the disk \citep{bovy2012,frinchaboy2013,bovy2014,nidever2014,Hayden_2015,Martig2015,anders2017, schiavon2017,Mackereth2017,hayes2018b,donor2018,Mackereth2019b,hasselquist2019,eilers2019,frankel2019,donor2020},
    the bulge \citep{nidever2015,ness2016,schultheis2014,Zasowski2016,Zasowski2019,Weinberg2019,Hasselquist2020,Rojas-Arriagada_2020,Griffith2020arXiv}, 
    and the halo \citep{fernandez_alvar_2017,fernandez_alvar_2018,hayes2018a,helmi2018,mackereth2019a,fernandez_alvar_2019,mackereth2020},
    as well as it's constituent globular clusters, satellite galaxies, and tidal streams \citep{majewski2013,meszaros2015,hasselquist_2017,schiavon2017,fu2018,hasselquist2019b,Masseron_2019,hayes_2020,horta2020a,meszaros2020}.

In addition to better understanding the Milky Way through classic Galactic astronomy exploration, as APOGEE-1 was originally designed to perform, the APOGEE dataset has also been used to study \added{scientific} fields outside of it's original intent, delving into
    stellar astrophysics \citep{apokasc,epstein_2014,apokasc2,Mackereth2020b},
    studies of the ISM \citep{schultheis2014,Zasowski_2015a,Zasowski_2015b},
    and even time series analyses of \replaced{rv}{radial velocity} variability 
        as an indicator of stellar companions \citep{troup_2016,badenes_2018,cunningham_2019,pricewhelan_2020,Mazzola2020}, 
        planet hosts \citep{Fleming_2015,canas_2018,wilson_2018,canas_2019,canas_2019a}
        and intrinsically variable stars \citep{chojnowski2015,Chojnowski2019,Chojnowski2020,Lewis_2020}.

This diversity of topics addressed is only possible because of the efforts put into the careful creation of the APOGEE data, which, on the one hand, needs to sample a broad range of stellar types and populations, but, on the other hand, must be chosen with simple enough criteria that a reliable selection function can be generated.
The selection criteria used to select stars for the different science programs in APOGEE-1 were presented in
\citet{zasowski2013}, which includes details on how these criteria have been optimized for the survey science goals, the tools required to calculate and account for the selection function, and how the different selection methods are identified using a set of targeting bits.

The second generation of the APOGEE project (APOGEE-2; S.~Majewski et al.~in prep.) is part of SDSS-IV \citep[2014-2020;][]{blanton2017}, and uses near-twin spectrographs \citep[described in][]{wilson2019} operating simultaneously in the Northern and Southern Hemispheres.
\added{The APOGEE-N spectrograph,} the original spectrograph used for APOGEE-1, remains mounted on the Sloan Foundation 2.5-meter telescope at the Apache Point Observatory \citep[APO;][]{gunn2006}, and is carrying out observations for the APOGEE-2 North (APOGEE-2N) program, while a second, near-clone spectrograph, APOGEE-S, \citep{wilson2019} is mounted on the Ir\'en\'ee~du~Pont 2.5-meter telescope \citep{bowen1973} at Las Campanas Observatory (LCO), where it is used to execute the APOGEE-2 South (APOGEE-2S) observing program. 

The expansion to the Southern Hemisphere in APOGEE-2 has provided the survey a truly panoramic view of the Milky Way, but, particularly, its “dust-hidden” inner regions.
The initial observational design of APOGEE-2 was explained in \citet[][hereafter Z17]{zasowski2017}, which included a detailed description for each of the distinct science programs planned for APOGEE-2 and was published alongside the first data release from APOGEE-2 \citep[in DR14][]{dr14,holtzman_2018}, \added{which contained only data from APOGEE-N.
We note that the \citetalias{zasowski2017} survey plan for APOGEE-2S was schematic in nature, as the APOGEE-S instrument was still being commissioned at the time of writing.}
It also included a general overview of a survey component unique to APOGEE-2S, that of Contributed Programs, observations allocated outside of SDSS-IV via proposals to use the APOGEE-S instrument through either the the Chilean National Time Allocation Committee (CNTAC) or the Time Allocation Committee (TAC) for the Observatories of the Carnegie Institution for Science (OCIS). 
However, \citetalias{zasowski2017} was written before survey observations on the du~Pont had commenced, and before the first Contributed Programs had even been scheduled\footnote{Time for Contributed Programs were allocated for observing semesters starting in 2017A and continued until 2020A.}. 
Thus, the first description of APOGEE-2S targeting \citepalias{zasowski2017}  both did not contain details for the Contributed Programs, and \deleted{a number of ensuing changes to the main survey targeting.} \added{also assumed that APOGEE-1 or APOGEE-2N strategies would largely be adopted identically for APOGEE-2S.}

\explain{ The following two paragraphs were entirely rewritten}
The main goal of this paper is to present the targeting strategy adopted for APOGEE-2S through the final stage of the APOGEE-2S operations that ended in January 2021.
The target selection methods for APOGEE-2S have evolved since the publication of \citetalias{zasowski2017}, and hence, could not have been contained in that or other past references.
Such changes were implemented after a detailed consideration of the early on-sky performance from APOGEE-2S \citep[DR16 was the first public release of data obtained with APOGEE-S spectrograph;][]{dr16,jonsson_2020}, in particular, the true observing efficiency and instrumental through-put.
With the experience and operational strategies of APOGEE-1 and APOGEE-2N in hand (see concluding section of \beatonetal), we were able to optimize the observations for APOGEE-2S in a timely manner.
We demonstrate how those modifications not only increase the likelihood of reaching our initial scientific goals, but also enabled us to exceed them in many cases.

Alongside this present paper, \beatonetal\ presents the final targeting strategy of APOGEE-2N, which is focused on its Bright Time eXtension and ancillary programs. 
Both papers are intended to complete the references about APOGEE target selection and complement the existing APOGEE targeting selection references \citepalias{zasowski2013,zasowski2017}, the data release references (\citealt{dr16}, \citealt{jonsson_2020}), and our online documentation (e.g., \url{https://www.sdss.org/dr16/} for DR16, \url{https://www.sdss.org/dr17/} for DR17). Together, these papers present the final field plan spatial coverage of APOGEE-2, as well as the statistics of our observing schedule. The papers focus on how the targeting strategy changed from that presented in \citetalias{zasowski2017}, explaining these differences for each scientific program separately, and also detailing the characteristics of any new programs. 
These articles also represent a unique opportunity to analyze how the differences between APOGEE-2S and APOGEE-2N, in terms of timeline of implementation, observing time availability per local sidereal time (LST), and telescope characteristics, impacted on the targeting strategy and observing progress for some programs.
In this paper dedicated to APOGEE-2S we also present descriptions for each of the programs contributed to the SDSS-IV survey from CIS and CNTAC allocations; the Contributed Programs represent $\sim$23\% of the total observing time for APOGEE-2S survey from 2017 to 2020 (For comparison, the APOGEE-2N Ancillary Programs were approximately 5\% of the 6-year program) and are a significant component of the APOGEE-2s dataset.

The outline of the paper is as follows: 
    Section \ref{sec:prelims} defines key concepts associated with \deleted{the survey} \added{SDSS-IV and APOGEE} that are relevant to targeting.
    Section \ref{sec:targoverview} gives an overview of how the different types of APOGEE-2S fibers (science, sky, and calibration) are assigned.
    Changes to the original field plan and targeting strategy for APOGEE-2S are given in Section \ref{sec:mainchanges} \added{and this section includes a table that summarizes the main scientific programs}. 
    Contributed Programs are detailed in Section \ref{sec:contributedprograms}. 
    A summary is given in Section \ref{sec:summary}. 
    A glossary of common terms used in this paper is presented in \autoref{glossary} to aid the reader.

\section{APOGEE Observations \& Targeting Basics} \label{sec:prelims}

\explain{ This entire section introduction was re-written}
In this section, we provide a summary of the primary methods used for selecting targets in APOGEE; much of this material is also presented in \citetalias{zasowski2013}, \citetalias{zasowski2017}, and \beatonetal, because these are the core set of principles guiding the design of all APOGEE observations.
Key SDSS- and APOGEE-specific concepts and terms affiliated with the process are described and defined; such terms are first introduced in quotation marks (e.g., ``term'') to indicate that they are specific technical terms. 
The key nomenclature are also organized as a Glossary in \autoref{glossary} as a convenience (following that of  \citetalias{zasowski2013} and \citetalias{zasowski2017}, but also including the new terms used for this present paper).
We also note that this section will include some detailed discussions on important topics that will be useful to guide the reader through the content of this paper.

Throughout this paper we will also refer to different ``tags'' that are part of various APOGEE-2 data products, and we will use their official names in those files to aid the user in finding data of interest. Each time we refer to one of those ``tags' we will use true-type fonts, (e.g., {\tt LOCATION\_ID}), and they will generally correspond to the ones present in the summary files \texttt{allStar} and \texttt{allVisit}, but they can also be found in several other APOGEE data products.

For reviewing the full description of the APOGEE data products, we refer the reader to our online documentation.\footnote{\url{https://www.sdss.org/dr16/irspec/spectro_data/}}

\subsection{Concepts and Nomenclature} \label{sec:concepts}

\explain{ For this subsection we changed the format from a running text to a definitions list following the referee's suggestion. We also changed the name of the section removing the word ``targeting'' from it.} 

\paragraph{Fields:}
\explain{ The following 2 paragraphs have been substantially changed}
The full observing program consists of a set of ``fields.'' 
A field is a particular location in the sky defined by a central coordinate and a radius that defines the field-of-view (FOV).
Targets are then selected for fields according to specific criteria that define where holes are to be drilled in corresponding fiber optic plugplates.

Each field is uniquely identified using an integer known as the ``location ID'' (in the data products, {\tt LOCATION\_ID}). 
We also assign a name to each field stored in the {\tt FIELD} tag that  follows the naming scheme of $LLL$+$BB$, where $LLL$ and $BB$ represent the central Galactic longitude and latitude rounded to the closest integer; for example, a field centered at Galactic coordinates $(\ell,b)=(44.76,-8.64)$ would have a field name of 045$-$09.
Some exceptions exist to this naming scheme for fields dedicated to specific astronomical objects like dwarf galaxies (e.g., with names like `CARINA' or `LMC1') or star clusters (like `N6441').
Fields that are close to each other on the sky may have the same name, but they will always have a unique location ID.

For Contributed Programs, a suffix is added to the \deleted{field} name that indicates the Time Allocation Committee (TAC) awarding the program. 
Fields from programs assigned through the CNTAC have a ``-C'' suffix, and fields from the OCIS TAC programs have an ``-O'' suffix \added{(e.g., 000+05-C or 001+02-O)}. 
A particularly large Contributed Program allocated through OCIS that focuses on NASA's Transiting Exoplanet Survey Satellite \citep[\tess,][]{ricker2015} Southern Continuous Viewing Zone have the suffix ``-O\_TESS'' (see \autoref{sec:teskevansaders}). 

\paragraph{Plates:}
To observe our targets in APOGEE-2S, we use standard SDSS plug-plates that hold the APOGEE-S fibers at the \added{sky location of the} desired target and a ``plate'' corresponds to the physical piece of metal used to hold the fibers at their sky positions, and each is identified with a plate ID.
When drilling a design of stars into a plate, atmospheric refraction corrections need to be taken into account to calculate the positions of the fibers on the plate and these corrections vary depending on the altitude and azimuth of the field over the night.
For this reason, a ``drill angle'' needs to be specified for each plate, which corresponds to the hour angle (HA) at which a plate is intended to be observed; 
the same set of targets can occupy plates drilled to different hour angles to increase the chances that that particular design will get observed.
Each plate is thus associated with a particular design and HA combination.

\paragraph{Design:}
Each APOGEE plate includes a set of stars for observation referred to as its ``design.''
While multiple plates can have stars in common, even a \deleted{single star of difference} \added{difference of a single star} implies that \deleted{ they are different designs and each is} \added{the design is different and would be} uniquely identified with its own design ID.\added{\footnote{The Design IDs are only found in Intermediate Data Products, see \url{https://www.sdss.org/dr16/irspec/spectro_data/}}}

A design is composed of one, two, or three ``cohorts'', which are sets of stars restricted to a specific magnitude range.
Stars are split into cohorts that receive different amounts of total exposure time such that all stars in a design will obtain roughly the same $S/N$ over the full range of magnitudes.
Bright stars are grouped into ``short cohorts'' (meaning that this group of stars would receive a relatively short amount of total exposure time via a smaller number of ``visits'' to those stars --- see below), medium brightness stars into ``medium cohorts'', and the faintest stars are grouped into ``long cohorts'' (receiving the largest amount of total exposure time).

Each design has a cohort ``version,'' which is a three digit number of the form $sml$ where $s$, $m$, and $l$ identify which of that field's short, medium, and long cohorts (uniquely numbered starting with 1), respectively, are included in that design.\added{\footnote{The cohort versions are only found in Intermediate Data Products, see \url{https://www.sdss.org/dr16/irspec/spectro_data/}}}
For example, a design for a given field that includes the second short cohort, the first medium cohort, and no long cohort would have cohort version $210$.
A design can, therefore, be thought of as a combination of a field and a cohort version. 
In their Figure 1, \citetalias{zasowski2013} provides a schematic example of how different cohorts are combined into different designs and distinct plates for a single APOGEE field.

\begin{figure}
    \centering
    \includegraphics[width=\columnwidth]{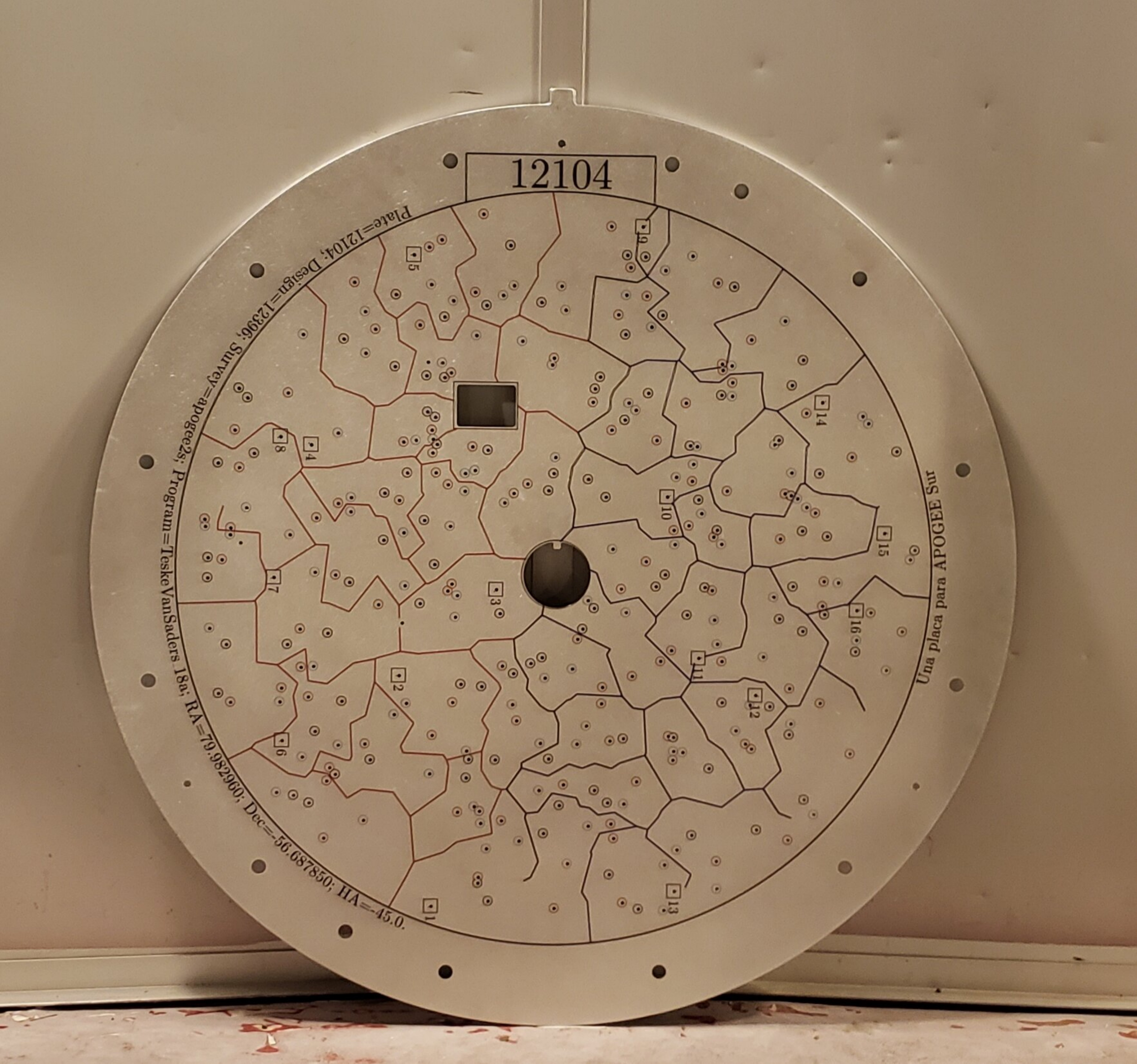}
    \caption{Example of an APOGEE-2S plate. 
    \deleted{Apart from covering} \added{APOGEE-S plates cover} a smaller area on the sky \added{than APOGEE-N plates} (due to telescope plate scale differences), \deleted{APOGEE-2S plates} \added{and they also} have a slightly different layout \deleted{than those from APOGEE-N} (compare to Figure 13 in \citealt{majewski2017}).
    Most obviously, there is a central hole for an on-axis acquisition camera and another rectangular hole for an off-axis camera. 
    These fiber exclusion regions are taken into account in the target selection and plate design software. 
    Another difference with APOGEE-N plates is that all plate markings, including coding of the holes to denote bright, medium and faint fibers, are automatically printed for the APOGEE-S plates, whereas they are hand-marked for the APOGEE-N plates \added{(the coding may not be visible without enlarging the image)}. 
    Photo by R.~Beaton.}
    \label{fig:apo2s_plate}
\end{figure}

\paragraph{Visits:}
The amount of time each star is observed is measured in \added{units of} ``visits'', where a single visit corresponds to a standard length of integration of $\sim$1~hour.  
Exposures are observed in pairs due to the need for half-pixel dithering in the spectral domain implemented for the recovery of Nyquist sampling of the native resolution delivered by the spectrograph optics onto the detectors (see \citealt{majewski2017,wilson2019}); in the Data Processing and Reduction pipeline \citep[DRP;][]{nidever2015}, these pairs of individually under-sampled exposures are effectively blended together via interpolation to achieve proper sampling of the resolution element \citep[for the DR16 implementation, see][]{jonsson_2020}.
In the end, a standard visit corresponds to a single sequence of observations comprised of eight exposures with a dither sequence of ABBA ABBA (where A and B refer to the two nominal dither positions), which sums to a total of 4000 seconds (the same as with APOGEE-2N).
In APOGEE-2, exceptions to this scheme were applied to those fields that contained five or more extremely faint stars ($H$ $\gtrsim$ 13.5). 
For these cases, the observing sequence was modified to enhance the exposure-level or visit-level $S/N$ for these targets in a particular visit.
Modifications could occur in two forms: 
    (i) ``DAB'' exposures, where only a single ABBA sequence occurs with each exposure at double length (for the same sum total exposure of 4000 seconds), thereby reducing the relative contribution of readnoise, and 
    (ii) ``TDAB'' exposures (for ``triple DAB''), where an extra, double length exposure AB pair is added to the sequence for a total visit length of 6000 seconds (ABBAAB).

\paragraph{Plate Design:}
The process of assigning targets to fibers takes into account the physical size of the fibers on the plate to avoid collisions between fibers and also considers regions in the focal plane used by each of the two acquisition cameras.
\autoref{fig:apo2s_plate} provides an example of an APOGEE-S plate. 
The ferrule that contains each fiber has a diameter equivalent to $56\arcsec$ in the focal plane and this physical size sets the lower limit of allowable separation between two fibers below which they will ``collide''  (\deleted{for context,} the fiber collision radius for APOGEE-N plates is 72$\arcsec$).
For initial field acquisition, a central and off-axis acquisition cameras (Manta G-235B) are employed; this is different to the setup for the Northern system, which relies upon coherent fiber bundles for field acquisition.\added{\footnote{The APOGEE-N plates have a central region of 96$''$ that cannot be used for targets due to a post that supports the plate, see \citet{Owen_1994}.}}
These cameras are used to zero-in on the central location of the pointing, determine the rotational alignment, and measure the focal plane scale.  
Use of these cameras creates regions that are not available for science fibers, and candidate targets at these positions are not permitted.
As shown in \autoref{fig:apo2s_plate}, these two acquisition cameras, which directly attach to the fiber plugplate, create exclusion zones in the center of the plate (with a circular footprint $5.5\arcmin$ in diameter) and somewhere off-axis that varies with each plate (with a rectangular footprint occupying an area equivalent to approximately $10\arcmin\times7\arcmin$).  
Though the physical plates in the South are the same physical size as the plates for APOGEE-N, due to differences in the telescope plate scales, the APOGEE-S plates cover a smaller area than APOGEE-N.
Ignoring the above exclusion zones, the maximal field-of-view of an APOGEE-S plate is 2.8~deg$^2$ (corresponding to a 1.9\degs\ diameter), compared to the 7.1~deg$^2$ FOV for APOGEE-N (corresponding to a 3.0\degs\ diameter) \citepalias[for a discussion of maximal FOV versus the drillable FOV, see section 2.1 of][]{zasowski2017}.  
To see other differences between the APOGEE-N and APOGEE-S plates, compare \autoref{fig:apo2s_plate} here to Figure 13 in \citet{majewski2017}.

\subsection{Targeting Bits} \label{sec:targbits}

Like APOGEE-1, APOGEE-2 uses target flags to indicate the reason why each target was selected for observation; the flags are encoded using bitmasks \added{in the data products}.
These flags are set to help reconstruct the target selection function of the survey.  
Note, however, that use of these flags may not allow for the complete retrieval of a desired set of stars.
As one example, stars selected for observation because they are red clump (RC) candidates have bit 25 set in \texttt{APOGEE2\_TARGET2}, but a
large number of RC stars are serendipitously included in our sample, and hence, will not have that bit set \deleted{(for a detailed discussion of APOGEE red clump stars see, e.g., Bovy et al. 2014)}.

The APOGEE-2 targeting bit labels are \texttt{APOGEE2\_TARGET1}, \texttt{APOGEE2\_TARGET2}, \texttt{APOGEE2\_TARGET3}, and, the recently defined, \texttt{APOGEE2\_TARGET4}.
Currently, \texttt{APOGEE2\_TARGET4} has no assigned bits, but we anticipate that it will be used for the final APOGEE-2 data release (DR17).
Table \ref{tab:targeting_bits} provides a summary of all APOGEE-2 targeting bits for both APOGEE-2N and APOGEE-2S. \added{In the following section, we provide brief descriptions of the major modifications to the Targeting Bits that were given in \citetalias{zasowski2017}.}

\begin{table*}[h]
\centering
\movetabledown=1.7in
\begin{rotatetable}
\caption{APOGEE-2 Targeting Bits \label{tab:targeting_bits}}
\footnotesize
\begin{tabular}{llllll} 
\hline \hline
\multicolumn{2}{c}{\texttt{APOGEE2\_TARGET1}} & \multicolumn{2}{c}{\texttt{APOGEE2\_TARGET2}} & \multicolumn{2}{c}{\texttt{APOGEE2\_TARGET3}} \\
{\it Bit} & {\it Criterion} & {\it Bit} & {\it Criterion} & {\it Bit} & {\it Criterion} \\
\hline \hline
0  & Single $(J-K_s)_0 > 0.5$ bin         & 0  & {\bf \ktwo\ GAP Program}                    & 0  & KOI target \\
1  & ``Blue'' $0.5 < (J-K_s)_0 < 0.8$ bin & 1  & \textbf{California Cloud Target}    & 1  & Eclipsing binary \\
2  & ``Red'' $(J-K_s)_0 > 0.8$ bin        & 2  & Abundance/parameters standard               & 2  & KOI control target \\
3  & Dereddened with RJCE/IRAC            & 3  & RV standard                                 & 3  & M dwarf\\
4  & Dereddened with RJCE/WISE            & 4  & Sky fiber                                   & 4  & Substellar companion search target \\
5  & Dereddened with SFD $E(B-V)$         & 5  & External survey calibration                 & 5  & Young cluster target \\
6  & No dereddening                       & 6  & Internal survey calibration (APOGEE-1+2)    & 6  & {\bf K2 Star} \\
7  & Washington+DDO51 giant               & 7  & {\bf Outer Disk Substructure Member}        & 7  & {\bf APOGEE2 Target} \\
8  & Washington+DDO51 dwarf               & 8  & {\bf Outer Disk Substructure Candidate}     & 8  & Ancillary target \\
9  & Probable (open) cluster member       & 9  & Telluric calibrator                         & 9  & {\bf Massive Star} \\
10 & {\bf Globular Cluster Candidate}     & 10 & Calibration cluster member                  & 10 & -- {\it QSOs} \\
11 & Short cohort (1--3 visits)           & 11 & {\bf K2 Planet Host}                        & 11 & -- {\it Cepheids} \\
12 & Medium cohort (3--6 visits)          & 12 & -- \textbf{\textit{Kepler Synchronized Binaries}}       & 12 & -- {\it The Distant Disk} \\
13 & Long cohort (12--24 visits)          & 13 & Literature calibration                      & 13 & -- {\it Emission Line Stars}\\
14 & Random sample member                 & 14 & Gaia-ESO overlap                            & 14 & -- {\it Moving Groups} \\
15 & MaNGA-led design                     & 15 & ARGOS overlap                               & 15 & -- {\it NGC 6791 Populations} \\
16 & Single $(J-K_s)_0 > 0.3$ bin         & 16 & {\it Gaia} overlap                          & 16 & -- {\it Cannon Calibrators} \\
17 & No Washington+DDO51 classification   & 17 & GALAH overlap                               & 17 & -- {\it Faint APOKASC Giants}\\
18 & Confirmed tidal stream member        & 18 & RAVE overlap                                & 18 & -- {\it W3-4-5 Star Forming Regions}\\
19 & Potential tidal stream member        & 19 & APOGEE-2S commissioning target              & 19 & -- {\it Massive Evolved Stars} \\
20 & Confirmed dSph member (non Sgr)      & 20 & {\bf Halo Member}                           & 20 & -- {\it Extinction Law} \\
21 & Potential dSph member (non Sgr)      & 21 & {\bf Halo Candidate}                        & 21 & -- {\it Kepler M Dwarfs} \\
22 & Confirmed Mag Cloud member 	      & 22 & 1-m target                                  & 22 & -- {\it AGB Stars} \\
23 & Potential Mag Cloud member 	      & 23 & Modified bright limit cohort ($H>10$)       & 23 & -- {\it M33 Clusters} \\
24 & RR Lyra star                         & 24 & Carnegie (CIS) program target               & 24 & -- {\it Ultracool Dwarfs} \\
25 & Potential bulge RC star              & 25 & Chilean (CNTAC) community target            & 25 & -- {\it SEGUE Giants} \\
26 & Sgr dSph member                      & 26 & Proprietary program target                  & 26 & -- {\it Cepheids} \\
27 & APOKASC ``giant'' sample 	          & 27 & {\bf N-CVZ OBAF stars}                      & 27 & -- {\it Kapteyn Field SA57} \\
28 & APOKASC ``dwarf'' sample 	          & 28 & {\bf N-CVZ GI Programs}                     & 28 & -- {\it K2 M Dwarfs} \\
29 & ``Faint'' target 			          & 29 & {\bf N-CVZ CTL star}                        & 29 & -- {\it RV Variables} \\
30 & APOKASC sample 			          & 30 & {\bf N-CVZ Giant with RPMJ}                 & 30 & -- {\it M31 Disk} \\
\hline 
\hline
\multicolumn{6}{l}{\added{Note 1: A new bitmask, \texttt{APOGEE2\_TARGET4}, has been added to the data model for DR17 but is currently unpopulated.}} \\ 
\multicolumn{6}{l}{Note 2: Flags that are new or different than what was presented in \citetalias{zasowski2017} are highlighted in bold.}
\end{tabular}
\end{rotatetable}
\end{table*}


\begin{figure*}[h]
    \begin{center}
    \includegraphics[angle=0, width=\textwidth]{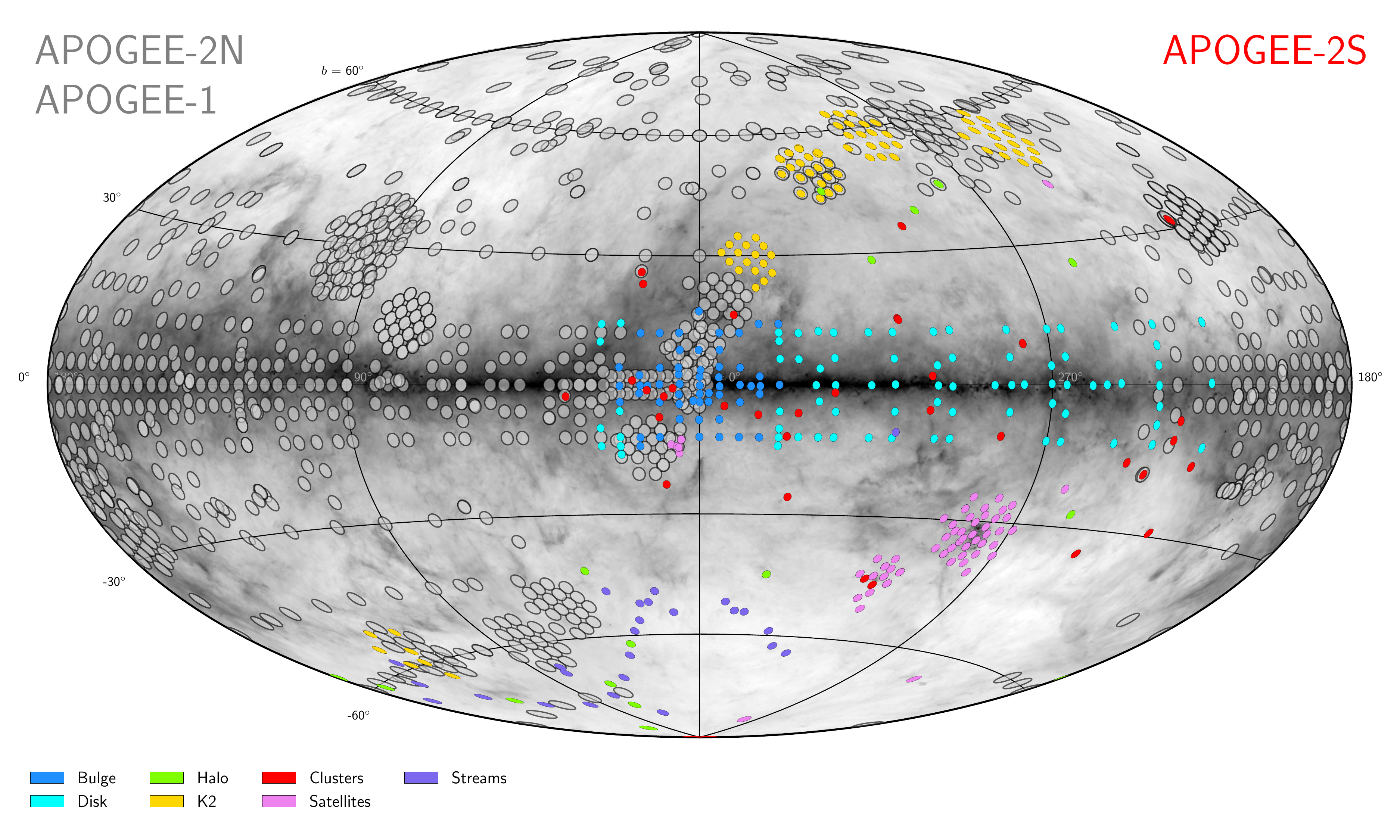}  
    \caption{\explain{ This figure and the caption have been updated} 
    Final APOGEE field plan in Galactic coordinates with APOGEE-2S fields from the main survey highlighted in color. 
    The color-coding corresponds to the scientific program affiliated with the field.
    The background image corresponds to the dust map from \citet{schlegel1998} and grey fields correspond to the APOGEE-1 and APOGEE-2N fields for context. 
    For completeness, we also show the other components of the APOGEE survey (APOGEE-1 and APOGEE-2N) as grey semi-transparent circles. 
    The meridians are spaced every $\Delta \ell$=90\degs\  and the parallels are spaced every $\Delta b$=30\degs\ .} 
    \label{fig:fieldmap}
    \end{center}
\end{figure*}

\subsection{Targeting Flag Changes} \label{sec:flag_changes}

In the middle of 2020, \deleted{we started a review of all the} \added{the APOGEE-2 team reviewed all} targeting flags to ensure the consistency of their application for APOGEE-2 targeting. 
\deleted{The review involved inspecting whether targeting flags were set as well as whether the definition of each target flag matched how it was employed.}
In the course of this process, we fixed some flags that were set incorrectly for certain classes of stars, added new targeting flags, and slightly modified some of the flag definitions to match better how they were used.

In Table \ref{tab:targeting_bits}, we present all of the \texttt{APOGEE2\_TARGET1}, \texttt{APOGEE2\_TARGET2}, and \texttt{APOGEE2\_TARGET3} flag bits and their descriptions.
\deleted{In that Table we also highlight, in bold face, all of the target flags that are new or different from what was presented in Z17} \added{All of the target flags that are new or different from what was presented in \citetalias{zasowski2017} are presented in bold face.} 

\deleted{In this subsection} \added{Below}, we list all the APOGEE-2 related targeting flags that were modified, either in their definition or in their implementation. 
Those flags that are highlighted in Table \ref{tab:targeting_bits}, but are not listed in this section are new flags that have been defined since \citetalias{zasowski2017} but were only used for APOGEE-2N, with their descriptions given in \beatonetal.
The one exception is \texttt{APOGEE2\_TARGET1}=6, which was used for \ktwo\ targeting and is described in Section \ref{sec:k2}.
\\

\noindent \textbf{APOGEE2\_TARGET1=29: ``Faint'' target.} \\
\explain{ This definition was rewritten entirely}
Even though this flag was defined prior to the review, it had only been used for a small group of targets for which it was set ``manually'' in plate design.
To make this flag consistent with its description, we systematically compared the $H$ magnitudes of stars with the number of visits they were planned to receive. We then flagged all stars that had fewer visits planned than what they needed to reach the nominal $S/N$ value lower limit of $100$ in Table \ref{tab:maglims}.
We highlight that there is a type of stars in out sample called ``special target'' (see \autoref{sec:specialtargs}) for which this flag was not set because the $S/N$ goals associated with those stars is, on occasion, different than the ones for the rest of our sample.\\

\noindent \textbf{APOGEE2\_TARGET2=16: \gaia\ overlap.} \\
This flag was previously designed to be used to flag any stars that overlapped with the \gaia\ \deleted{sample. Because} \added{observations. Given that} no stars were intentionally targeted because of their overlap with \deleted{the \gaia\ sample, this flag} \added{\gaia\,, this flag, as implemented,} did not serve to identify a reason for why specific stars were targeted. 
\added{Moreover,} given that \gaia\ is photometrically-complete well beyond APOGEE-2's faintest magnitude limits, such a flag would nominally include nearly all of APOGEE-2's targets.  
In addition, a \gaia\ \deleted{DR2} cross-match is provided with APOGEE-2 data products \cite[e.g., \added{for DR16 a cross match to \gaia\ DR2;}][and for DR17 a crossmatch with EDR3, Holtzman et al. in prep.]{jonsson_2020}, providing a clearer way to identify that stars have overlap between \gaia\ and APOGEE.  
Since overlap with \gaia\ could be easily ascertained from other information provided by APOGEE-2, this flag, therefore, had less meaning than originally intended.

The meaning of this targeting flag has been changed to identify any targets that explicitly used \gaia\ data (almost exclusively proper motions) for the target selection process, and we have set this flag for those stars. \\

\noindent \textbf{APOGEE2\_TARGET1=10: Globular Cluster Candidate.} \\
\deleted{This is} A new flag used for globular cluster candidates that were selected either by photometry only or by membership probabilities based on proper motions. At the time of their selection no such flag existed, so targets selected as Globular Cluster candidates were not identified in any way\deleted{and there was no way to determine the reason for their selection}. This has now been remedied. \\

\noindent \textbf{APOGEE2\_TARGET2=0: \ktwo\ GAP Program.} \\
This is a new flag defined to identify targets selected because they were part the \ktwo\ Galactic Archaeology Program (GAP) \citep{Stello_2017}.  Most of these stars were already flagged with \texttt{APOGEE2\_TARGET3=6}, which is used for all \ktwo\ targets, but now all \ktwo\ targets, additionally, have flags associated with their \deleted{\ktwo\ subprogram (see Subsection 4.7)} \added{corresponding priority group, which are listed in \autoref{sec:k2}}. \\

\noindent \textbf{APOGEE2\_TARGET2=20: Halo Members.} \\
This flag was \deleted{already used for halo targets but previously was specifically used for stars} \added{previously used to identify halo targets specifically} selected as part of the APOGEE-2N Bright Time eXtension (BTX). Now we have extended the meaning of this flag to include a similar class of stars in APOGEE-2S that were selected using information from the SkyMapper survey \citep{keller2007}, \deleted{as discussed in} \added{see} \autoref{sec:halo}. \\

\noindent \textbf{APOGEE2\_TARGET2=21: Halo Candidates.} \\
\deleted{As with the previous flag, this flag was also} \added{This flag was} originally defined to be used for stars from the APOGEE-2N BTX, but \deleted{it's meaning has now expanded to incorporate} \added{it now also includes} halo candidates from APOGEE-2S that were selected using \gaia\ proper motion information\deleted{ as described in} (\autoref{sec:halo}). \\

\section{Targeting Selection Overview} \label{sec:targoverview}

For APOGEE-2, the $300$ fibers of each APOGEE spectrograph are divided among $250$ science targets, $35$ sky targets, and $15$ telluric targets.\footnote{APOGEE-1 used $230$ fibers for science targets, $35$ for sky targets, and $35$ for telluric targets. See \citetalias{zasowski2017} for the motivations for this change.}
This section provides an overview of how each of these targets types are selected.

\subsection{Telluric Absorption Calibration Targets} \label{sec:tells}

The wavelength range of the APOGEE spectrographs contains a number of contaminant spectral features from the Earth's atmosphere \deleted{, such as CO$_2$, H$_2$O, and CH$_4$ absorption bands and OH airglow emission lines}.
To measure the telluric absorption we need to observe stars whose spectra are as close as possible to a featureless blackbody\deleted{; given that hot stars are the best approximation to this ideal, we select the bluest stars}\added{, and thus, we select the stars with the bluest observed $J-K_{\rm s}$ colors} in each field for this purpose. 
The spectra of these stars are processed by the APOGEE data reduction pipeline to make the telluric corrections \citep[apred;][]{nidever2015,holtzman_2018,jonsson_2020}.
\deleted{The telluric absorption calibrators are selected as the stars with the bluest observed $J-K_{\rm s}$ colors; thus, these targets are not de-reddenned before selection as are the science targets.}
To take into account angular variations in the telluric absorption during observations, telluric calibrators are spatially distributed across the plugplate as homogeneously as possible.
For this purpose, the FOV for each field is divided into a number of segmented, equal-area zones (Figure 8 of \citetalias{zasowski2013} provides a zone schematic) \added{such that a telluric calibrator can be selected for each zone}.
\deleted{The first half of telluric calibrators are chosen by selecting the star with the bluest color in each zone. The second half of the calibrator sample is composed using the bluest stars remaining across the entire field, regardless of zone.}
These \replaced{tellurics}{stars} are the first ones selected for APOGEE-2S plates and can be identified with bit 9 in \texttt{APOGEE2\_TARGET2}.

\subsection{Sky Contamination Calibration Targets} \label{sec:skies}
\explain{ This subsection was rewritten entirely}
The APOGEE data reduction pipeline uses observations of  ``empty'' sky regions to monitor the airglow and other foreground and background emissions.
To select sky targets in a field, we first select candidate positions that correspond to locations that have no 2MASS sources within a $6\arcsec$ radius.
Then, as we do for selecting telluric absorption calibrators (\autoref{sec:tells}), the plugplate FOV is split into equal area zones (Figure 8 of \citetalias{zasowski2013} provides a zone schematic) and we select up to eight candidate positions for each zone, to create the final list of sky positions.
These fibers are the third and last type of targets selected in APOGEE plates.
Sky positions will have bit 4 set in \texttt{APOGEE2\_TARGET2}, but typically these spectra are not kept throughout the reduction process.

\subsection{Science Target Selection} \label{sec:science}

Our science programs are designed to address many distinct scientific goals, which can require targeting specific Galactic sub-components or specific stellar types.
\deleted{To address these goals, we have to select the optimal sample of stars for each field by applying different selection criteria using observed photometric parameters.}
The subsections that follow explain the most important aspects of the science target selection process.

\added{In the process of filling an APOGEE plate with the different classes of targets,} science targets are selected after telluric correction targets but before sky targets are selected.
There are two major types of science targets in APOGEE, the ``main red star'' sample, which is randomly selected to map the bulge, disk, or halo based on \deleted{photometric parameters} \added{their apparent magnitude and color values}, and ``special targets'', which correspond to specific high priority targets (e.g., red clump, stream members, RR Lyr \deleted{type} stars).
In this latter class, targets can be selected using complex combinations of photometric, chemical, and kinematical information. 

``Special targets'' have top priority and then the ``main red star'' sample is the source for all remaining fibers.
In the following subsections, we explain how these two types of stars are selected.

\subsubsection{Main Red Star Sample} \label{sec:maintargs}

The ``main red star'' sample comprises the majority of the APOGEE-2S stars by number. 
The underlying targeting strategy for this sample \deleted{was the use of} is a simple color-magnitude criterion to select stars from the bulge, disk, and halo (these are given in \autoref{tab:colorcuts}).
\deleted{Targets from ``main red star'' sample are included in fields targeted by all APOGEE-2S programs.}

To select targets for the ``main red star'' sample, we start with all of the objects in the 2MASS Point Source Catalog \citep[PSC;][]{skrutskie2006} that fall in the FOV of a given field. 
The NIR photometry is complemented by mid-IR photometry from either the \spitzer$+$IRAC GLIMPSE \citep{benjamin2005,churchwell2009} or the AllWise \citep{wright2010} \deleted{, after using}. The
Rayleigh-Jeans Color Excess method (RJCE) \added{is then used} to estimate the line-of-sight extinction for each individual target \citep[][also discussed below]{majewski2011,zasowski2013,zasowski2017}.
Data quality limits are applied to ensure that the sources have small magnitude uncertainties and reliable quality photometry flags \citepalias[for the specifics, we refer the reader to Table 2 of][]{zasowski2017}.

\begin{table}[h]
    \centering
    \caption{Cohort Visits and Magnitude Limits}
    \label{tab:maglims}
    \begin{tabular}{ccc}
        \hline \hline
        NVisits & Hmin  & Hmax \\
                & [mag] & [mag]\\
        \hline 
        1  & 7.0  & 11.0 \\
        3  & 11.0 & 12.2 \\
        6  & 12.2 & 12.8 \\
        12 & 12.8 & 13.3 \\
        24 & 13.3 & 13.8 \\
        \hline 
        \hline
    \end{tabular}
\end{table}

To maximize the number of giant stars targeted while ensuring a simple selection function, we apply a simple color selection using the de-reddened $(J-K_{\rm s})_0$ color.
For most APOGEE fields, we use the RJCE method to estimate $E(J-K_{\rm s})$ \citep{majewski2011}; RJCE assumes that, for filters sampling the Rayleigh-Jeans 
tail of stellar spectra, all stars nominally have the same color such that any deviation from that baseline color (in this case we adopt the  $(H-I2)$ or $(H-W2)$ indices) is due to foreground dust.
\added{RJCE is used for the main red star sample in nearly all disk and bulge fields.}

RJCE is known to be less reliable when the total $E(J-K_{\rm s})$ is of order the color-uncertainty in a given field; \deleted{(e.g., for low extinction fields)} \added{more specifically, for fields with low extinction that are well out of the Galactic Plane (e.g., the typical uncertainty on $(H-W2)$ is larger than the extinction in these fields)}. 
To avoid the overestimation of $E(J-K_{\rm s})$ for low extinction halo fields, the reddening was determined using the dust maps from \citet{schlegel1998}.

\explain{ The next paragraph is new}
Because multiple dereddening methods are used, the targeting flags indicate the method adopted for a given target.
More specifically, the dereddening method used for each target is indicated by the \texttt{APOGEE2\_TARGET1} flags recorded in:
    bit $3$ for RJCE using \spitzer$+$IRAC, 
    bit $4$ for RJCE using \emph{WISE}, 
    bit $5$ using \citet{schlegel1998}, or 
    bit $6$ if no de-redenning was applied to that target.
In addition to the targeting flags, the summary files contain a number of convenience tags that provide the values used for extinction; these include {\tt AK\_TARG} for the $A_{K}$ used in targeting, {\tt AK\_TARG\_METHOD} with a string defining the method, {\tt AK\_WISE} with the all sky RJCE value from WISE photometry, and {\tt SFD\_EBV} with the all sky value from \citet{schlegel1998}\footnote{For additional description, please see the data model here: \url{https://data.sdss.org/datamodel/files/APOGEE_ASPCAP/APRED_VERS/ASPCAP_VERS/allStar.html }}. 
The appropriate 2MASS, WISE and \spitzer\ photometry, when available, is also provided in the summary files.

The color criteria for selecting red stars in a given field depends on the Galactic component being targeted; these criteria are summarized in \autoref{tab:colorcuts}.
The color selection strategy employed for each target is indicated in the \texttt{APOGEE2\_TARGET1} flag by 
    bit 0 for \deleted{a single} $(J-K_{\rm s})_0$ \textgreater\ 0.5, 
    bit 1  for the ``blue'' bin of $0.5~\leq~(J-K_s)_0$~\textless~0.8, 
    bit 2  for the ``red'' bin of $(J-K_{\rm s})_0$ \textgreater\ 0.8, and, 
    bit 16 for \deleted{a single bin of} $(J-K_{\rm s})_0$ \textgreater\ 0.3. 

\begin{table*} 
    \centering
    \caption{Color Cuts for Galactic Regions \label{tab:colorcuts}}
    \begin{tabular}{c cc l l l} 
    \hline \hline
    Galactic & $\ell$ & $b$   & Color Selection$^{a}$ &  \multicolumn{2}{l}{Targeting Flag in {\tt APOGEE2\_TARGET1}$^{b}$} \\
    Region   &  Range & Range & [mag]           &      bit    &  Description \\
    \hline 
    Bulge  & \textless~20\degs or \textgreater~340\degs & \textless~25\degs & $0.5~\leq~(J-K_{\rm s})_0$         &  0 & APOGEE2\_ONEBIN\_GT\_0\_5 \\
    Disk   & $\geq$~20\degs and $\leq$~340\degs         & \textless~25\degs & $0.5~\leq~(J-K_s)_0$~\textless~0.8 &  1 & APOGEE2\_TWOBIN\_0\_5\_TO\_0\_8 \\
           &                                                &                     & $0.8~\leq~(J-K_{\rm s})_0$         &  2 & APOGEE2\_TWOBIN\_GT\_0\_8 \\ 
    Halo   & no $\ell$ limits                               & $\geq$~25\degs   & $0.3~\leq~(J-K_{\rm s})_0$         & 16 & APOGEE2\_ONEBIN\_GT\_0\_3 \\
    \hline \hline
    \multicolumn{6}{l}{ (a) The values for a star are coded in the {\tt MIN\_JK} and {\tt MAX\_JK} tags.} \\
    \multicolumn{6}{l}{ (b) The equivalent bit for APOGEE1 is {\tt APOGEE1\_TARGET1} and it follows the same definitions.}
    \end{tabular}
\end{table*} 


For some fields targeting the halo ($|b|~\geq$~25\degs), we apply an additional photometric criterion designed to distinguish between dwarf and giant \deleted{type} stars of the same spectral type. 
This classification is performed by combining the Washington filter system with the $DDO51$ filter (hereafter W+D), an intermediate-band filter that spans the surface gravity-sensitive Mgb feature in the spectra of cool stars,
to construct a  color-color diagram, $(M-T_2)$ against $(M-DDO51)$. 
Because of the gravity sensitivity of the $DDO51$ filter \citep{McClure_1973} and $T_{\rm eff}$ sensitivity of the Washington filters \citep[here $M$ and $T2$][]{Canterna_1976}, this color-color space provides an effective means of selecting likely giants even in dense foregrounds \citep[e.g.,][]{geisler1984,majewski2000,munoz2005}.
\citetalias{zasowski2013} (\citetalias{zasowski2017}) provides a detailed discussion of how this photometry is used for APOGEE-1 (APOGEE-2) targeting. 
This same methodology is used for the APOGEE-2S targeting when W+D photometry was available (the photometry and its application are discussed in \citetalias{zasowski2017}). 
When assigning priorities to the targeting in halo fields that have W+D photometry, the first priority were W+D selected giants, then stars with no W+D classification, and lastly, \deleted{only if open fibers remained,} W+D classified dwarfs were used to fill remaining fibers.

Finally, to attain a goal $S/N$ of $100$ per pixel in the spectra, magnitude limits must be set on the targets that are selected for each design. 
Because each field design has short, medium, or long cohorts that receive different numbers of visits, the magnitude limits applied to the cohorts are modified according to the number of visits that each cohort will receive (following \autoref{tab:maglims}).  
\deleted{For reference,} The total number of visits for a field is typically equivalent to the number of visits planned for the longest cohort (i.e., faintest target) included in its designs.

\subsubsection{Special Targets} \label{sec:specialtargs}

Some APOGEE-2S fields are designed with special scientific goals in mind and, therefore, include a 
list of specific stars to be targeted at the highest priority, which we refer to broadly as ``special targets.''
Typically, ``special targets'' are provided by a specific Science Working Group in prioritized lists \deleted{(e.g., the Open Cluster Working Group, see Donor et al. 2020)}. 
In some cases, the ``special targets'' for an APOGEE-2S field may use all of the available fibers, as in the Large Magellanic Cloud program \deleted{(Nidever et al. 2020; Zasowski et al. 2017)}, while other programs only require a handful of fibers, as in the Open Cluster program.
\deleted{Thus, after these targets are selected,} Any remaining, unallocated fibers are used to bolster the ``main red star'' sample. \deleted{as described in Subsubsection 3.3.1, following the color-magnitude selection appropriate for the primary targeted Galactic component and the number of visits.}

The Working Group for each APOGEE-2S program sets their own criteria to select high priority targets; Table \ref{tab:programs} provides a brief summary of these programs. \deleted{as well as the appropriate reference to check for details about the selection process of ``special targets'', when applicable.}
Details about the original selection method used for all APOGEE-2S programs is described in \citetalias{zasowski2017}.
Stars that were selected as different types of ``special targets'' can also be identified according to their set targeting bits (\autoref{tab:targeting_bits}).

\begin{table*}
\centering
\caption{Summary of APOGEE-2S Main Survey Programs
\label{tab:programs}}
\scriptsize
\begin{tabular}{p{2.2cm} p{0.7cm} p{0.7cm} p{0.8cm} p{1.8cm} p{4.5cm} p{2.7cm} p{1.8cm}}
\hline \hline
Name & Total Fields & Total Visits & Section & Special Target Flags & Systems Included & Additional References (if applicable) & \texttt{PROGRAMNAME} \explain{New}\\
\hline 
Bulge             & $56$ & $347$ & \ref{sec:bulge} & --- & --- & \citetalias{zasowski2017} & bulge\\
Disk              & $78$ & $579$ & --- & --- & --- & \citetalias{zasowski2017} & disk, disk1, disk2\\
Halo              & $15$ & $126$ & \ref{sec:halo} & Target1=18 Target2=20,21 & \added{Halo stars,} Sagittarius stream, Orphan stream & \citetalias{zasowski2017} for main red star sample; \citet{hayes2018b} for Sagittarius stream targets & halo, halo2\_stream\\
Streams           &  $26$ &  $132$ & \ref{sec:streams} & Target1=18,26 Target2=20,21 & Jhelum stream, \added{ Saggitarius stream,} Orphan stream & --- & sgr\_tidal, stream\_halo, stream\_disk\\
Open Clusters     & $15$ & $87$  & \ref{sec:openclusters} & Target1=9 & Berkeley~75, Berkeley~81, M~8, M~16, M~67, NGC~2204, NGC~2243, NGC~6253, NGC~5999, NGC~6583, NGC~6603, Trumpler~20, Trumpler~32, Tombaugh~2, Collinder~261 & \citet{donor2018} & cluster\_oc\\
Globular Clusters & $20$ & $147$ & \ref{sec:globularclusters} & Target1=10 Target2=2,10 & 47~Tucanae, M~10, M~12, M~22, M~4, M~55, M~68, M~79, NGC~1851, NGC~2808, NGC~288, NGC~2298, NGC~3201, NGC~362, NGC~6388, NGC~6397, NGC~6441, NGC~6752, Omega Centauri & --- & cluster\_gc\\
Magellanic Clouds & $48$ & $483$ & \ref{sec:magclouds} & Target1=22,23 & LMC, SMC & \citet{nidever2019a} & magclouds\\
Dwarf Spheroidals &  $4$ &  $96$ & \ref{sec:fornax} & Target1=20,21 & Carina, Sextans, Sculptor , Fornax & \citetalias{zasowski2017} & halo\_dsph\\
Sagittarius       &  $6$ &  $36$ & --- & Target1=26 Target2=10 & Sagittarius dwarf core \deleted{and stream} & --- & sgr \\
\ktwo\            & $87$ &  $87$ & \ref{sec:k2} & Target1=30 Target2=0,11 Target3=6,28 & K2 Campaigns: C6, C8, C10, C14, C15, C17 & \beatonetal & k2 \\
\hline 
\end{tabular}
\end{table*}

\section{Changes to the Main Survey}
\label{sec:mainchanges}

\explain{ This paragraph is new}
In this section we detail all the changes made to
the targeting strategy, which includes the target selection and field locations of APOGEE-2S, with respect to the original
plan given in \citetalias{zasowski2017}.
Subsection \ref{sec:motivation} provides the motivation for these 
changes and, in particular, contains an overview of the time-allocation for APOGEE-2S.
After this overview, the subsections that follow are organized by APOGEE-2S science program and describe the changes applied to the targeting 
strategy.

\subsection{Motivation} \label{sec:motivation}

One of the main motivations for this paper is to describe modifications to the original APOGEE-2S survey strategy described in \citetalias{zasowski2017}.
These modifications occurred for three reasons: 
    (1) internal evaluations via Targeting Reviews\added{, using our ``Lessons Learned'' from APOGEE-N,} of whether the survey strategies in place would enable APOGEE-2S to achieve its scientific goals, 
    (2) the advent of new data from surveys outside of SDSS (e.g., {\it Gaia}) that could significantly enhance APOGEE targeting efficiency, and
    (3) modifications to the main survey plan due to alterations in the observing schedule. 
    
Internal evaluations of all APOGEE-2 programs occurred on an \deleted{regular} approximately yearly basis. 
For some scientific programs, the review suggested that a modification to the original targeting strategy was needed to meet the science goals, or that the program would significantly benefit from considering additional aspects for the target selection process.
In other cases, the sky-efficiency was either lower or higher than originally estimated; this particularly occurred at some local sidereal time (LST) windows for which the weather was better than our adopted weather model.

Additional motivations for modifying the targeting strategy came by way of previously unavailable information from
space-based missions like 
    \gaia\ \citep{gaia,gaia-dr1,gaia-dr2},
    the \ktwo\ mission of the \kepler\ spacecraft \citep{howell2014},
    or \tess\ \citep{ricker2015}, 
as well as complementary ground-based surveys like
    GALAH \citep{zucker2012}, 
    SkyMapper \citep{keller2007}, or 
    RAVE \citep{steinmetz2006,kunder2017}.

Finally, modifications to the APOGEE-2S observing schedule occurred both because of the addition of some nights to the originally expected allocation or shifts in the scheduled time\added{. This schedule was then modified for two main reasons:} a delay in commissioning \deleted{at the beginning of the survey} \added{from August 2016 to October 2016} and a forced and lengthy closure of Las Campanas Observatory due to the COVID19 pandemic near the end of the survey. 
The total observing hours and their distribution throughout the year determine the total number of visits we expect to observe at each LST, and thus, are crucial for estimating the \deleted{distribution of} fields that we will be able to observe.
All of these factors were considered simultaneously to design and implement modifications to what became a somewhat dynamic targeting scheme, but one that always pointed at maximizing scientific return.

In the end, a total of 352 nights \replaced{are expected}{were allocated} for the Main Survey programs\footnote{With 10 nights in 2017 used for the commissioning of the APOGEE-S spectrograph, which are not counted toward the Main Survey observing total.} \deleted{ and an additional 104 nights to be used for Contributed Programs following the ratio of Main Survey versus Contributed Program nights of approximately 3:1 per annum} \added{from April 2017 to January 2021}.
More details about the allocation of du Pont 2.5-m time to the APOGEE project will be given in the forthcoming APOGEE-2 overview paper (S.~Majewski et al.~in prep.). 

The final field map for the APOGEE-2S Survey is given in \autoref{fig:fieldmap}, where we show the positions, in Galactic coordinates, for all the Main Survey APOGEE-2S fields.
The fields in APOGEE-2S are color-coded by their science program.
We also show in grey circles the APOGEE-1 and APOGEE-2N fields to exhibit the overall coverage of the APOGEE survey. 
This figure shows how APOGEE-2S makes APOGEE a truly panoramic survey reaching all regions and components of the Milky Way by covering regions inaccessible from the Northern Hemisphere, most notably the inner Galactic regions.

\subsubsection{Observing efficiency of APOGEE-2S and resulting modification to visit plan \label{sec:allocations_efficiency}}

\explain{The title of this subsection was changed}
The observing efficiency of \deleted{APOGEE-2 is} \added{APOGEE-2S was} continuously calculated \added{during survey operations} and compared to the initial estimate \added{used for the \citetalias{zasowski2017} plan}.
The original observing plan of the survey was constructed assuming a seasonally averaged observing efficiency of $\sim$4.7 visits per night, and involved $1493$ visits planned to be observed over the entire \added{APOGEE-2S} survey period from \deleted{2016 to 2020} \added{February 2017 to June 2020}. 

During the first year of the survey (2017), operational strategies were under development and initial plate delivery timescales were sometimes mismatched to the optimal observation dates.
This translated in miscalculations of the visits expected per LST for the first plateruns, which in turn resulted in nights that had some unused visits because there were no suitable plates.
Additionally, confronted with a completely new instrument and still evolving observing infrastructure that had many significant differences with that in place for decades at APO, 
the APOGEE-2S engineering and observing teams took some time to optimize operational strategies and 
software and to become proficient in their use.
Together, these factors translated to a lower than expected efficiency of $\sim$3.1 visits per night during that year (65\% of the anticipated efficiency).
These initial challenges, however, were solved and, for \deleted{2017} \added{2018} and the first half of \deleted{2018} \added{2019}, the observing efficiency grew to $\sim$5.8 and $\sim$5.5 visits per night, respectively (123\% and 117\% of the anticipated efficiency).

Assuming that the observing efficiency would remain constant until the end of the survey, we re-estimated that our overall observing efficiency would be \replaced{$\sim$~4.9}{$\sim$~5.0} visits per night.
Using this new efficiency, we concluded that in our total \deleted{$352$} APOGEE-2S nights we would be able to observe $\sim224$ visits more than our original expectations.
For this reason, the original field plan was modified from a $1493$ visit plan, to one containing $1717$ total visits. 
From these extra visits $58$ (26\%) were necessary to account for the modifications made to the Bulge and RRL Programs, while the others were used to observe new fields that were not in the original targeting plan.

\subsubsection{COVID-19 Shutdown and Final Plan} \label{sec:covid19}

Due to the COVID-19 pandemic's global spread in early 2020, LCO shut down all operations in mid-March 2020.  
This included APOGEE-2S, which had to stop operations about 6 months before the nominal conclusion of the APOGEE-2S survey.  Ultimately, this resulted in a loss of 77 originally planned APOGEE-2S nights \deleted{($\sim$377 visits)}.
When LCO reopened in early October 2020, APOGEE-2S was granted an extension that encompasses the original 77 nights lost due to the COVID-related closure as well as 11 additional nights. 

The total APOGEE-2S extension period includes 88 nights over the October 20, 2020 - January 21, 2020 time frame \deleted{($\sim$430 visits)}.
While this extension allowed APOGEE-2S to (more than) recover its lost time, the allocated nights gave very little access to the same portion of the sky, which necessitated a major overhaul of the plan for the remainder of the survey, with modified objectives to make compelling use of the altered sky access.
The most notable impact on the survey was a loss of almost the entire final year (i.e., about 1/3) of the planned APOGEE observations of the central Milky Way.

The COVID-19 extension left a projected \replaced{350}{403} visits to fill with newly drilled plates, but a short timeline in which to produce them, and with the imminent danger that any resumption of operations might be tentative.
Under such circumstances, it was decided that the most expeditious path forward was to expand those APOGEE-2S programs already primarily focused on those regions of the sky --- namely the Magellanic Cloud and the Galactic Halo Programs, both of which would not only benefit from greater coverage, but also could be flexible in the event of additional closures or extensions.
With the extra \replaced{350}{403} visits from the COVID-19 extension the APOGEE-2S observing plan reached its definitive version, consisting of \replaced{2067}{2120} visits, \added{spread over 352 nights,} whose sky coverage is shown in Figure \ref{fig:fieldmap}.
In the end, despite numerous obstacles the survey had overcome, the size of the final observing plan represents an increase of \replaced{$\sim{35\%}$}{$\sim{42\%}$} with respect to the original $1493$-visit plan, albeit with some shift in emphasis from that originally envisioned.

The following sections present all of the changes to the APOGEE-2S observations originally planned and presented in \citetalias{zasowski2017}. \added{The presentation is} organized by scientific program.

\noindent\begin{table*}
    \centering
    \caption{APOGEE-2S Galactic Bulge Observing Plans \label{tab:bulge}}
    \scriptsize
    \begin{tabular}{|cccc|cc|cc|cc|}
        \hline \hline
        & & & & \multicolumn{2}{c|}{``Short'' Cohort} & \multicolumn{2}{c|}{``Medium'' Cohort} & \multicolumn{2}{c|}{``Long'' Cohort} \\
    Plan     & Stars & Visits & Fields & \multicolumn{2}{c|}{$H<11.0$ Stars} & \multicolumn{2}{c|}{$11.0<H<12.2$ Stars} & \multicolumn{2}{c|}{$12.2<H<12.8$ Stars} \\
     & & &        
     & $S/N>100$   &     $80<S/N<100$
     & $S/N>100$   &     $80<S/N<100$
     & $S/N>100$   &     $80<S/N<100$ \\
    \hline
    Original & 33374 & 264 & 61 & 22751 & 0 & 10223 & 0    & 400  & 0 \\
    New      & 27168 & 347 & 56 & 10355 & 0 & 6525  & 3732 & 6253 & 303 \\
    Percent Change & -19\% & +31\% & -9\% & -55\% & 0\% & -37\% &  & +1463\% & \\
    \hline
    \end{tabular}
\end{table*}


\begin{figure*}
    \begin{center}
    \includegraphics[angle=0, width= \textwidth]{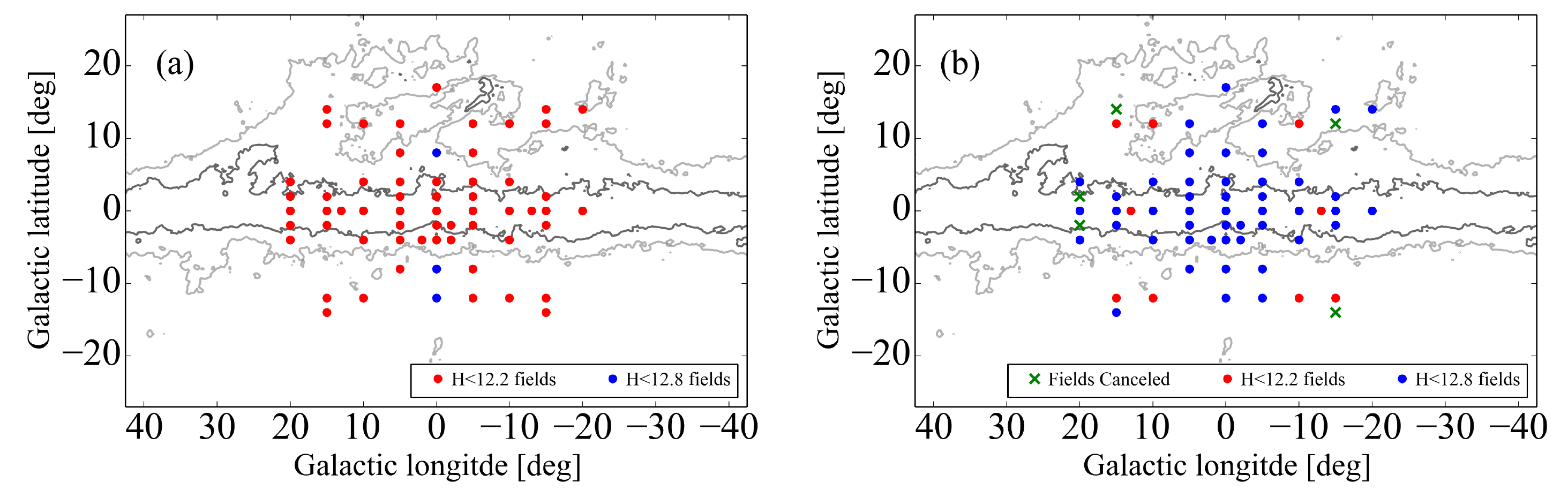}  
    \caption{\explain{ This figure has been updated}The APOGEE-2S coverage and depth plan for the Galactic bulge. 
    The full Galactic bulge region is shown with medium (red) and long (blue) fields indicated. \added{Grey contours show color excess $E(B-V)$ levels of of 0.5 (inner line) and 1.5 (outer line).}
    \replaced{(Left)}{(a)} The original plan, which was composed of 3-visit fields to a depth of $H\sim$12.2 (red), with three deeper pointings at $\ell$=0\degs (blue). 
    \replaced{(Right)}{(b)} The final plan, which demonstrates the increase in fields reaching $H\sim$12.8 (blue), the elimination of some fields (black crosses), as well as the fields that were properly designed for $H\sim$12.2 that were unchanged.
    As discussed in the text, the result of these modifications to the survey is that our mapping of the Galactic bulge goes deeper and has built a larger sample of bone-fide bulge stars \citep[see e.g., ][]{Rojas-Arriagada_2020,Hasselquist2020}.
    }
    \label{fig:bulgemap}
    \end{center}
\end{figure*}

\subsection{The Milky Way Bulge} \label{sec:bulge}

\deleted{We discovered} In May 2018\added{, when the second of our four total bulge seasons was starting we discovered}  that, due to an error in the execution of the targeting selection software, a significant fraction of the plates in the Bulge Program contained stars fainter than the originally intended magnitude limit for the plate (e.g., see \autoref{tab:maglims}).
In this subsection, we will briefly explain the origin of this problem, the actions taken to resolve it, and how the implemented solution, along with an increased observing efficiency, fortuitously resulted in a deeper, more complete survey of the Galactic bulge than originally planned.

\subsubsection{The Problem}

The spatial distribution of the fields in the Bulge Program is given in \autoref{fig:bulgemap}, which shows the original plan (\autoref{fig:bulgemap}a) and the final plan (\autoref{fig:bulgemap}b).
The color-coding in \autoref{fig:bulgemap} indicates the targeting depth of either $H$=12.2 ($S/N$=100 in 3-visits; red) or $H$=12.8 ($S/N$=100 in 6-visits; blue). 
The intended Bulge Program (\autoref{fig:bulgemap}a; \citetalias{zasowski2017}) was to sample the bulk of the Galactic bulge region in a grid with plates going to $H$=12.2 depth, with a handful of deeper fields at strategic locations designed to complement the APOGEE-1 Bulge Program \citepalias{zasowski2013}.
However, the Bulge Program that was implemented in the target selection software is shown in \deleted{the right panel of}  \autoref{fig:bulgemap}b; the latter reflects the fact that the majority of the bulge fields were inadvertently drilled with stars as faint as $H$=12.8 (blue) and, given that observations had already begun, the options to resolve the problem were limited. 
Were the nominal survey minimum $S/N$ requirement enforced, completion of the observing plan shown in \deleted{the right panel of} \autoref{fig:bulgemap}b would have required about twice as much time as was available for the Bulge Program.

More specifically, the original Bulge Program (\autoref{fig:bulgemap}a) contained \replaced{32}{34} fields meant to be observed with 3-plates each.
Each of the fields included three short cohorts scheduled for 1-visit each and a single medium cohort scheduled for 3 visits (in the cohort nomenclature introduced earlier, a ``310 plate'').
Ten of the bulge fields were intended to have 6-visits total, but in order to observe more of the stars in these dense bulge fields, they were intended to have a 3-visit depth like the above fields, having two pairs of 3-visit medium cohorts and six 1-visit short cohorts (e.g., a 620 cohort complement, with two sets of 310). 
In both the 3-visit and 6-visit fields, the 1-visit short cohorts were erroneously filled with stars as faint as $H$=12.2 (i.e., what is normally a 3-visit depth) and some of the 3-visit cohorts were filled with stars as faint as $H$=12.8 (i.e., the normal 6-visit depth).  
This is problematic for APOGEE observations, because, in 1-visit, the faintest stars in the short cohorts would be expected to achieve $S/N\sim58$ and, in 3-visits, the faintest stars in the medium cohorts should reach $S/N\sim75$ --- i.e., neither cohort satisfying the APOGEE goal of $S/N$=100 for all targets.\footnote{For context, the ASPCAP pipeline automatically flags anything with $S/N<70$ with \texttt{SN\_WARN} or bit 11 of \texttt{ASPCAPFLAG} \citep[see Table 11 of ][]{holtzman2015}.}

While the fields just described formed the bulk of the bulge design problem, there were other bulge fields where the magnitude limits were properly set given the scheduled number of visits, and a few fields where the number of visits was larger \added{than} needed \added{to reach $S/N$=100}. However, \deleted{on average, the field plan that was erroneously created would have taken considerably more observing time to achieve nominal $S/N$ of 100 than we available, and a mitigation strategy was needed.} \added{the overall bulge field plan was erroneously created, because achieving $S/N$ of 100 in all these fields would have required considerably more observing time than what we had available. Therefore, we needed a mitigation strategy.}

\subsubsection{The Solution}
By the time this situation was discovered, \added{the first out of our four total bulge seasons was complete and we had already designed and drilled all the plates for the second season. This means that by that moment} a significant fraction of the problematical plates had already been drilled, and some were even \deleted{in the process of being} observed.
For that reason, re-designing, re-drilling, and re-shipping these plates with the aim of restoring the original plan was deemed impractical, and a solution was sought that optimized utilizing the plates already in hand.
This required finding a compromise between $S/N$ needs and other modifications that made optimal use of the available, bulge-accessible observing nights for APOGEE-2S.

The solution that was devised involved the following action items:
\begin{itemize}
  \item Cancel all pending visits to plates that would yield $S/N > 100$ for the faintest stars on the plate \deleted{(although this was a problem for only a few fields)}. 
  The three fields involved correspond to those indicated by blue circles in \autoref{fig:bulgemap}a and \autoref{fig:bulgemap}b.
  \item Cancel visits for the second set of three plates for the 6-plate fields (i.e., remove the second 310 cohort set from a 620 cohort complement field). 
  Because there are no common stars between the two sets of medium cohorts (and all 1-visit cohorts are distinct), canceling one set does not affect the $S/N$ of the stars in the other set. 
  This change maintains our spatial coverage, but with a smaller total sample in those fields.
  \item Cancel fields from bulge regions that had another field nearby and/or a field symmetric with respect to the $\ell=0$\degs\ axis. 
  These \added{five} fields are indicated with an `x' marker in \autoref{fig:bulgemap}b.
  \item Increase the number of total visits for cohorts whose faintest stars are $H\sim12.2$ stars from 1-visit to 2 visits. 
  This change ensures that all targets reach $S/N>\sim80$ \added{These stars have been flagged as ``Faint'' targets with targeting flag APOGEE2\_TARGET1=29}.
  \item Increase the number of total visits for cohorts whose faintest stars have $H\sim12.8$ from 3 visits to 6 visits.
  This change ensures that all stars in these fields reach $S/N>$100. 
\end{itemize}
\noindent Implementation of the above action items required a net increase in the total number of visits assigned to the Bulge Program from 264 to 347 visits (a 31\% increase).

\deleted{To help make up this 82 visit difference required making an additional, difficult compromise:} \added{To find this time,}
we elected to cancel the 25 Main Survey visits allocated to the Bulge RR Lyrae (RRL) Program described in \citetalias[][their section 4.9]{zasowski2017}.
At the time that the Bulge Program was redesigned, a Contributed Program with the same goals (\autoref{sec:kollmeier1}) had already completed a number of visits that far exceeded the 25-visits initially scheduled as part of the original APOGEE-2S targeting plan.
Since the Contributed Program data become part of the overall APOGEE-2 dataset, the Science Requirement goal for the Main Survey RRL program ($\sim$4000 stars) had formally been met (albeit not through the route intended), whereas, at the time, the ``main red star'' sample from the Bulge Program were not meeting their Science Requirement.

\deleted{Even with this sacrifice,} \added{Still,} 58 more visits had to be found to complete the new observation plan for the Bulge Program.
\deleted{However,} At that point in the survey (October 2018), based on evidence for a growing APOGEE-2S observational efficiency that would\deleted{, in the end,} deliver a \added{final} mean value that exceeded the estimate used in the design of the original survey (as discussed in Section \ref{sec:motivation}), we found that the remaining \replaced{57}{58} visits would be made up through faster completion of plates overall.

\subsubsection{Summary of Bulge Modifications}

Table \ref{tab:bulge} provides a summary of the most relevant information about the intended and actual Bulge Programs, in which we can see that even though the total number of bulge targets has lessened from the original plan by $\sim19\%$ ($6206$~stars), the stars ``lost'' in the modified plan are the brightest, with $H<11.0$; typically, stars brighter than $H\sim11$ are more likely to be disk stars in the foreground to the bulge.
The new plan significantly increases the number of stars in the magnitude range $12.2<H<12.8$, which were practically non-existent in the original plan \added{, and the overall sky coverage of the new plan (\autoref{fig:bulgemap}b) is almost equal to the one of the original plan (\autoref{fig:bulgemap}a)}.
It is also worth mentioning that even though the new plan will produce $\sim$15\% of stars observed in the Bulge Program with $S/N < 100$ per pixel, these stars will obtain, at a minimum, $S/N \sim 80$ per pixel, which does not represent a considerable cost to our science goals given current pipeline performance \replaced{\citep{jonsson_2020}}{\citep{jonsson_2020,majewski2017}}.
\deleted{This is because the vast majority of the stellar parameters derived from those spectra will still reach APOGEE's precision requirements (see Majewski et al. 2017 for details).}
\deleted{Figure 3 shows that the new bulge targeting plan (right panel) has essentially the same sky coverage as the original plan (left panel).
The largest difference is that the vast majority of the fields in the new plan now contain targets to $H=12.8$ (i.e., almost twice as faint as originally intended and more homogeneously distributed across the Galactic Bulge footprint), and all stars will have a goal $S/N > 80$.}

\subsection{Galactic Halo and Stellar Streams} \label{sec:halostream}

The observing scheme for the Galactic halo and halo substructure \deleted{(e.g., tidal streams)} in APOGEE-2 (both APOGEE-2N and APOGEE-2S) changed significantly from that presented in \citetalias{zasowski2017}.
The original Halo Program \deleted{described in Z17} was a uniform grid of pointings out of the Galactic plane, covering a range in $|b|$ from $30$ to $75$ degrees.
While visits were set aside to sample streams in APOGEE-2N, no stellar stream observations were planned for APOGEE-2S, with the exception of a single field targeting the Sagittarius stellar stream (field name SRGT-2). 

In APOGEE-2S, only two of the original halo grid fields, 280+45 and 320+45, were \deleted{initially designed with 24 visits planned for each} \added{designed, each with 24 visits planned}.
The targets for these halo fields were selected using a color criterion of $(J-K_{\rm s})_0 > 0.3$ (dereddened color and employing the W+D giant method, \autoref{sec:maintargs}) and no other targeting criterion. 
\deleted{The vast majority of final fields in the Halo Program} \added{All the other fields from the Halo Program}, however, were designed using a new targeting strategy \added{created to maximize the number of halo- or stream- candidates that could be observed with the LST-visits available}.

\explain{The following paragraph was rewritten entirely}
Then, the final APOGEE-2S Halo Program totals 41 fields and 258 visits: 48 visits split between the original 280+45 and 320+45 dedicated halo fields, 78 visits divided among the 13 new dedicated halo fields, 12 visits in the original dedicated stream field SGRT-2, and 120 visits split between the 25 new dedicated stream fields.
The specific targeting priorities for each of these programs and their specific fields are given in the next subsections. 

Dedicated halo fields have \texttt{PROGRAMNAME} tag value ``halo''\footnote{There is one exception for the field \texttt{066-79} that has \texttt{PROGRAMNAME} ``halo2\_stream'' due to an error.} and their field names were assigned based on their Galactic coordinates like all disk and bulge fields.
For the dedicated stream fields, Sgr stream fields have \texttt{PROGRAMNAME} ``sgr\_tidal,'' and other stream fields have \texttt{PROGRAMNAME} of ``stream\_halo'' or ``stream\_disk,'' depending on the Galactic coordinates, and the field names for this program indicate the stellar stream being targeted. \added{Even though the 15 dedicated halo fields and the 26 dedicated stream fields are both part of the APOGEE-2S Halo Program, we show their statistics in \autoref{tab:programs} and their spatial distribution in \autoref{fig:fieldmap} as separate cases to provide more level of detail for the reader.}

\subsubsection{More Efficient Halo Targeting} \label{sec:halo}
In this section we explain \deleted{the different selection methods used in APOGEE-2S for halo fields} \added{how halo targets were selected for the fields in our Halo Program (while the selection of stream targets in these fields is explained in \ref{sec:streams})}.
An analysis of the data from APOGEE-2N halo observations suggested that the target selection was not observing enough stars at large line-of-sight distances to place them securely in the outer halo.
This process and resulting changes are described in our companion paper on APOGEE-2N targeting \beatonetal, but we summarize the key results here.

Prior to the Bright Time Extension (BTX), the Halo Program for APOGEE-2N relied on ``deep'' and ``narrow-area'' targeting relying on the Washington$+DDO51$ photometry technique to identify likely giant stars from the dwarf star foreground \citepalias[see][]{zasowski2013,zasowski2017}. 
While this strategy worked at $\sim$80\% efficiency to identify giants, the method was not particularly effective at identifying stars whose distances placed them securely in the outer halo; 
specific heliocentric distance limits were made in the APOGEE-2 Science Requirements Document for a certain number of halo stars to be surveyed at $>$15~kpc and $>$25~kpc, whereas the actual yields were around 3$\times$ smaller than anticipated for survey goals (\beatonetalpar).
By studying the proper motions of distant stars identified in APOGEE-1 and APOGEE-2N, it was found that proper motions could be used to remove the vast majority of the foreground dwarf contamination, which, in turn, greatly increased the likelihood of targeting a rare distant giant in a given field. 
As shown in \beatonetal, the distant star yields using this procedure were between 2-3 times more effective than the original selection procedure.
Because the APOGEE-2S halo targeting had not yet been fully implemented when the investigations described in \beatonetal\ were concluded, it was possible to incorporate the results of that investigation \textit{en masse} for the APOGEE-2S Halo Program.

\explain{ All the (originally 8) paragraphs from this point to the end of the subsection have been rewritten. The information presented is the same but the changes in order and wording are so many that is easier to read without highlights}
To maximize the number of halo stars in our APOGEE-2S Halo Program, we selected targets using the following priority ranking:

\begin{enumerate} \itemsep -2pt
\item The highest halo candidate/member priority targets were selected from the SkyMapper survey \citep{keller2007}. 
\citet{casagrande2019} used SkyMapper ($uvgriz$) and 2MASS ($JHK$) photometry to produce $T_{\rm eff, phot}$ estimates at a precision of $\sim$100~K, metallicity estimates at 0.2~dex precision (for [Fe/H]$>-2$), and a reliable dwarf/giant separation. 
Using these data,\footnote{The catalog is available: \url{https://github.com/casaluca/SkyMapper}.} halo candidates were identified using the following criteria: 
[Fe/H]$_{\rm phot}$\textless$-0.9$ and 
3200~K $<$ $T_{\rm eff, phot}$ $<$ 5500~K.
Stars targeted using these photometric stellar parameters have bit 20 set in \texttt{APOGEE2\_TARGET2}. This targets were included in all the fields from our Halo Program excepting the two original dedicated halo fields 280+45 and 320+45.
\item Then, we selected proper motion candidates taken from the \gaia~DR2 \citep{gaia-dr2}. 
The criteria for selecting these stars was $\mu$ \textless~5\,mas yr$^{-1}$ and $\sigma_{\rm {\mu}} /\mu$\textless~0.1 for all the 13 new dedicated halo fields, and \textless~5\,mas yr$^{-1}$ and $\sigma_{\rm {\mu}}$\textless~0.2 for the 18 new sgr\_tidal stream dedicated fields. For the 15 fields sgr\_tidal4 to sgr\_tidal18 a magnitude limit of $H < 13.3$ was used for selecting these stars, while a magnitude limit of $H < 12.8$ was used for all other fields, which is standard for 6-visits fields (see \autoref{tab:maglims}).
Stars selected using these proper motion restrictions have bit 21 set in \texttt{APOGEE2\_TARGET2} flag.
\item Then, for all APOGEE-2S Halo program fields containing W+D photometry we incorporated W+D photometric giant candidates (see \autoref{sec:maintargs}) and these targets have bit 7 set in \texttt{APOGEE2\_TARGET1}.
\item Finally, all APOGEE-2S Halo program fields that still have remaining unassigned fibers are back-filled with stars from the ``main red star'' sample following the magnitude and color ranges appropriate for the number of visits, as given in Tables \ref{tab:maglims} and \ref{tab:colorcuts}, respectively.
\end{enumerate}
\begin{figure*}
    \centering
    \includegraphics[width=\textwidth]{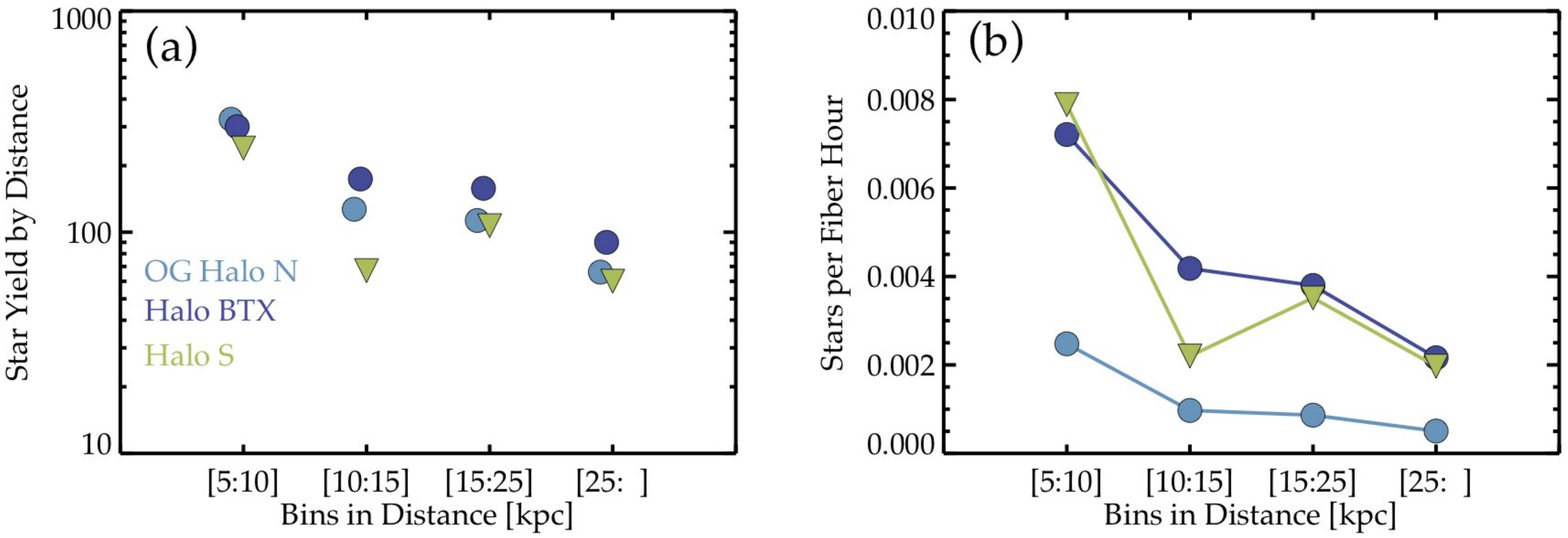}
    \caption{ \explain{ This figure has been updated}
    Comparison of the targeting efficiency for giants in Halo Programs. 
        We compare the original halo targeting from APOGEE-2N that relied on Washington$+DDO51$ (``OG Halo N;'' blue circles), the BTX APOGEE-2N targeting (``Halo BTX;'' purple circles), and the APOGEE-2S halo targeting (``Halo S;'' green triangles).
        For these figures, we only consider those stars that have measured stellar parameters (\logg\  and \teff) and we utilize spectrophotometric distances following the analysis described in \citet[][and references therein]{rojas-arrigada_2020}. 
        (a) Number of stars in different distance bins. 
        (b) Overall efficiency at identifying distant halo stars normalized by the total fiber hours in the program. 
        The total fiber hours are computed by summing the number of visits that contribute to the final spectrum used for ASPCAP analyses for all of the targets in the program (\texttt{NVISITS}).
        }
    \label{fig:halo_efficiency}
\end{figure*}

In Figure \ref{fig:halo_efficiency}, we show the efficiency of the different halo targeting strategies used throughout APOGEE-2.
\deleted{as of an internal data release of APOGEE data through 2020-03-24. In Figure \ref{fig:halo_efficiency}, we compare three different methods that the APOGEE-2 survey has used to select halo targets:}
\added{These methods are:}
    (i) the original method used for APOGEE-2N based on W+D photometry for giant-dwarf pre-classification (labelled ``OG Halo N''),
    (ii) the method used in the APOGEE-2N BTX based on HSOY proper motion selection \citep{altmann2017} and SEGUE-confirmed distant K-giants (labelled ``Halo BTX'') \citep{xue2014}, and
    (iii) the modified APOGEE-2S halo selection described here, based on SkyMapper data and proper motion information (labelled ``Halo S'').

The left panel of Figure \ref{fig:halo_efficiency} shows the number of stars from each Halo Program obtained within broad distance bins. 
Overall, the total number of stars in each distance bin is more-or-less consistent between the three targeting methods, but to compare how successful each method was we also need to consider the observing time invested in each Halo Program.
Thus, to calculate the efficiency of each method, we normalized the number of stars observed within each distance bin by the total number of ``fiber hours'' from each program, where a \textit{fiber hour} is one visit for a single fiber from a plate, and a single plate visit has a total of 265 fiber hours dedicated to stars. \added{This means that a single star from the Halo program observed N times adds N fiber hours to total of the Halo program, and for plates dedicating fibers to more than one program (e.g., APOGEE-2N plates with both Halo and ancillary program stars) only the fibers dedicated to the Halo program would count towards the total sum of fiber Hours for the Halo.}

The right panel of Figure \ref{fig:halo_efficiency} shows the number of Halo Program stars obtained per distance bin divided by the total number of fiber hours associated with each Halo Program. 
The APOGEE-2S strategy is about 3 times more effective than the original APOGEE-2N halo strategy in terms of observing distant halo stars for all distance bins. 
This demonstrates that altering the targeting strategy for APOGEE-2S has resulted in a much more efficient method for pre-selecting distant halo stars than originally used in APOGEE-2. 
On the other hand, the APOGEE-2S and APOGEE-2N BTX halo targeting strategies show similar efficiencies, as displayed in Figure \ref{fig:halo_efficiency}b. 
Achieving this for APOGEE-2S is remarkable given that APOGEE-2N includes known halo members from SEGUE which were not available for APOGEE-2S, and the plates in APOGEE-2N have a larger FOV that enable mapping larger region of the sky in a single visit.

\subsubsection{Targeting Stellar Streams} \label{sec:streams}
\explain{ The first paragraphs of this subsection and the Sagittarius stream selection section have been rewritten entirely. The Sagittarius stream selection method sections have been merged into a single Sagittarius stream selection section to increase clarity.}
To increase the number of halo stars in APOGEE-2S while simultaneously sampling halo substructure, we added 25 dedicated fields, whose locations were placed on the footprints of known stellar streams.
This represents one of the most important changes to our observing plan, considering that originally the Halo Program only contained one dedicated stream field called SGRT-2.
The 26 dedicated stream fields were designed targeting three streams in the Southern Hemisphere, and were split in the following way: 
\begin{enumerate}
    \item 19 fields placed on the Saggitarius (Sgr) stream, consisting of field SGRT-2, fields Sgr\_tidal1 to Sgr\_tidal3, and fields sgr\_tidal4 to sgr\_tidal18. All these fields have \texttt{PROGRAMNAME} ``sgr\_tidal''
    \item 6 fields (JHelum1 to JHelum6) placed on the JHelum stream. All these fields have \texttt{PROGRAMNAME} ``stream\_halo''
    \item 1 field placed on the Orphan stream (Orphan1), with \texttt{PROGRAMNAME} ``stream\_disk''
\end{enumerate}

Besides these 26 dedicated stream fields, Sgr stream targets were added to 8 dedicated halo fields (032-62, 037-43, 060-72, 066-79, 120-85, 137-71, 173-61, 179-57), and Orphan stream targets were added to the dedicated halo field 339-44. This means that our Halo Program includes a total of 27 fields targeting Sgr stream targets, 6 fields targeting JHelum stream targets, and 2 fields targeting Orphan stream targets. Throughout this subsection we are going to explain how we selected targets from these 3 streams in all these fields.\\

\noindent \textbf{\textit{Sagittarius Stream Selection:}}~
The original dedicated stream field SGRT-2 selected targets from the Sgr stream based on W+D dwarf/giant classification, and the location of the stars in a  $([J-K_{\rm s}]_0,H)$ CMD \citepalias[see][for more details]{zasowski2017}.
In the new APOGEE-2S Halo Program, we included Sgr stream targets in 18 new dedicated stream fields, and 8 new halo dedicated halo fields, by using the candidate stars presented in \citet{hayes2018c} as our parent sample.

To describe the regions observed in this stream, we use the heliocentric Sgr spherical coordinate system presented in \citet{majewski2003}. In this system, the equator is defined by the Sgr debris midplane and corresponds to a pole with Galactic coordinates $(\ell,b)=(273.8,-13.5)$. Then, Sgr latitudes, $|B_{\odot}|$, are defined as the angular distance to the Sgr debris equator as viewed from the Sun and Sgr longitudes, $\Lambda$, are defined to increase in the direction of trailing Sgr debris, with $\Lambda$=0 defined as the longitude of Sgr center from the King profile fitting results of \citep{majewski2003}.

\citet{hayes2018c} identified candidate Sgr stream red giants in the magnitude range of $10 < H < 13.5$ lying at Sgr stream latitudes of $|B_{\odot}| \lesssim$ 20\degs\ along $\sim$90\degs\ of the trailing arm of the Sgr stream. Because the Sgr stream is kinematically colder than the rest of the local halo, these candidates could be selected confidently based on their proper motions, particularly in having small, halo-like proper motions that are oriented in the direction of motion of the Sgr stream \citep[see][]{hayes2018c}.
The locations for the 18 new stream dedicated Sgr tidal fields (called Sgr\_tidal1 to Sgr\_tidal3, and sgr\_tidal4 to sgr\_tidal18) were selected to maximize the number of Sgr stream candidates from this sample and the spatial coverage along the stream.
The 8 new dedicated halo fields also included Sgr stream candidates from the Sgr sample, but the field location was chosen to maximize halo targets (see \autoref{sec:halo}) over stream targets. 
As a result of these changes, the APOGEE-2 coverage of the Sgr stream goes more or less homogeneously from $\Lambda$ = 35\degs\  to $\Lambda$ = 110\degs\ at Sgr stream latitudes of $|B_{\odot}| \lesssim$ 10\degs\ ($\ell$ from 10\degs\ to 175\degs\ and $b$ from -48\degs\ to 82\degs).\\
 \\

\noindent \textbf{\textit{Jhelum Stellar Stream }}\added{ \textit{Selection}}: ~ Six \added{dedicated} stream fields were designed to target the Jhelum stream \citep{Bonaca:2019,schipp2018}, with \texttt{FIELD} of ``JHelum1'' through ``JHelum6''.
The fields were selected by tracing the stream \replaced{in}{using} \gaia~DR2 photometry and astrometry.
Candidates were selected as high-probability members of the stream based on their proximity (in sky position) to the stream track defined by \citet{Bonaca:2019}, and by modeling the (solar motion corrected) proper motion distribution along the stream.

In detail, following the method used in \citet{Price-Whelan:2018}, we construct a mixture model for the full proper motion distribution in this part of the sky.
A Gaussian mixture model (GMM) representation of the ``background'' proper motion distribution was first constructed by fitting a GMM to the proper motion distribution of stars in a sky track offset by two degrees north and south (in a Jhelum stream-aligned coordinate system; \citealt{Bonaca:2019}).
At locations along the Jhelum stream track, we then fit a mixture model containing the GMM background model as one component, and a single Gaussian with varied mean and variance as the ``stream'' component. 
We also fit for the mixture weight, which defines the relative contribution of the stream and background components to the total proper motion distribution.

This model is then used to assign membership probabilities to stars in the vicinity of the Jhelum stream.
Stars were selected with $>50\%$ membership probability that also lie within 0.25\,mag of the red giant branch for a a metal-poor ($[{\rm Fe}/{\rm H}]=-2$) isochrone (from the MIST stellar isochrone library; \citealt{Dotter:2016, Choi:2016}). By this means, a total of seven candidate stars from the Jhelum stream were selected. 
The dedicated publication on the Jhelum stream provides some additional details \citet{Sheffield2021arXiv}. \\

\noindent \textbf{\textit{Orphan Stellar Stream}}\added{ \textit{Selection}}:~ Orphan stream \citep{belokurov2006, Koposov:2019}\deleted{, whose} targets were included in the \added{dedicated} halo field 339-44 and the \added{dedicated} stream field Orphan1. 
To select candidate Orphan stream members, we use the sample of RR Lyrae-type Orphan stream stars 
from \citet{Koposov:2019} to fit polynomial trends in sky position and the two proper motion components (all in the Orphan stream coordinate system defined in \citealt{Koposov:2019}).
We then select all sources with \gaia~DR2 astrometry and 2MASS photometry in a 70\degs\  by 10\degs\  region around the extension of the Orphan stream into the Southern Hemisphere. 
We use the polynomial tracks of the stream \deleted{(determined from the RR Lyrae star members)} combined with a color-magnitude selection using a metal-poor ($[{\rm Fe}/{\rm H}]=-2$) isochrone (from the MIST stellar isochrone library; \citealt{Dotter:2016, Choi:2016}) to select candidate red giant branch star members of the stream.
A dedicated publication on the Orphan stream will provide additional details (K.~Hawkins et al.~in prep.).

All stars targeted as stellar stream candidates have bit 18 set in \texttt{APOGEE2\_TARGET1}.
The remaining targets in each field were selected following the halo targeting strategy discussed in \autoref{sec:halo}.

\subsection{Dwarf Spheroidals: Fornax} \label{sec:fornax}
\explain{ The title of this subsection has changed. Original name was Fornax dSph}
Because of improvements in survey efficiency and the allocation of extra nights to APOGEE-2S it was possible to expand the coverage of the survey and add an additional dwarf galaxy to APOGEE's sample, the Fornax dSph galaxy \added{\citep{Shapley_1938}}. 
\deleted{Fornax (Shapley 1938), is so distant  ($m-M=20.84\pm0.18$ or $D_{\rm gal}=147\pm12$\,kpc; McConnachie 2012), that the tip of the red giant branch lies at $H\sim$14.7, where, at the outset of APOGEE-2, the data performance at these faint magnitudes was not well-established, hence it's exclusion from the original targeting plan; however, results from the dSph program with APOGEE-2S have been encouraging and the \gaia~DR2 proper motions provide an improved foreground removal strategy.}
As for all of the other dSphs in APOGEE-2S, Fornax is planned to be observed with a single cohort of targets for 24 visits. 
Stars for this field were selected using a combination of spectroscopic, photometric, and proper motion information that we now describe.

At highest priority, we targeted four of the five confirmed Fornax globular clusters \citep{Hodge_1961}, three of which were targeted as integrated light observations (Fornax 3, 4, and 5), while for the fourth cluster (Fornax~2) we specifically targeted the brightest star.
These integrated light observations complement programs in APOGEE-1 for M\,31 globular clusters \citep[][Z13]{Sakari_2016} and in APOGEE-2N for M\,33 clusters (a description is given in \beatonetalpar).

At second priority we targeted individual Fornax member stars identified by prior radial velocity and chemical abundance analyses using VLT$+$FLAMES from \citet{Letarte_2010} and \citet{Lemasle_2014}. 
Additional radial velocity members from the wide-area survey with Magellan$+$M2FS by \citet{Walker_2009} were added at third priority.

Because many spectroscopic studies have small footprints compared to the APOGEE-2S FOV, additional candidate members were selected by combining ground-based photometry with \gaia~DR2 proper motions. 
The photometry came from two sources: 
    (i) optical CFHT photometry from \citet{Munoz_2018} (highest priority) and 
    (ii) NIR photometry from 2MASS \citep{skrutskie2006}.
Lastly, we included photometric candidates from the NIR study of \citet{Gullieuszik_2007}.
In all categories, stars below the tip of the red giant branch (i.e., with magnitudes $14.7\,< H < 15.2$) were selected ahead of stars with asymptotic giant branch (AGB) like magnitudes above the tip of the red giant branch (i.e., $H < 14.7$).

\subsection{Magellanic Clouds} \label{sec:magclouds}

Targeting of the Magellanic System --- the Large and Small Magellanic Clouds (LMC, SMC) and the substructure in their vicinity --- formed a major component of the APOGEE-2S program; a summary is given in \citetalias{zasowski2017} with additional details and the first results of the program given in \citet{nidever2019a}. 
For reference, the Magellanic Clouds program pulled targets from a variety of targeting classes to cover a wide range of luminous stellar evolutionary states, while also building a sample of red giants, which could be reliably analyzed by APOGEE and compared to Milky Way red giant samples.  
\deleted{These} \added{For context, the} targeting classes for general LMC and SMC targeting are as follows, in order of priority:
\begin{enumerate} \itemsep -2pt
    \item Supergiants following \citet{Neugent_2012} and \citet{Bonanos_2009} (limited to $\leq20$ per plate), 
    \item Hot main-sequence stars (limited to $\leq20$ per plate), 
    \item \citet{Olsen_2011} retrograde stars (limited to $\leq20$ per plate),  
    \item Post-AGB stars from \citet{Kamath_2014,Kamath_2015} (limited to $\leq10$ per plate),
    \item RGB stars with high resolution public spectroscopy (limited to $\leq10$ per plate), 
    \item AGB Carbon-rich stars following \citet{Nikolaev_2000} (limited to $\leq20$ per plate), 
    \item AGB Oxygen-rich stars (limited to $\leq20$ per plate), 
    \item RGB stars ($\geq$130 per plate).
\end{enumerate}
\deleted{An expanded description and visualization of the selection criteria can be found in Nidever et al. (2020).Magellanic Cloud members have targeting bit 22 set in \texttt{APOGEE2\_TARGET1}, while Magellanic Cloud photometric candidates have targeting bit 23 set in \texttt{APOGEE2\_TARGET1}.}
All designs belonging to Magellanic Cloud fields in the Main Survey have \texttt{PROGRAMNAME} tag value of ``magclouds.''

The Magellanic Clouds Program was augmented in three ways: 
    (1) expanded LMC and SMC spatial coverage, 
    (2) a optical cross-calibration field, and
    (3) the inclusion of a Contributed Program in the Main Survey. 

\noindent \textbf{\textit{Expanded Spatial Coverage:}}~ 
The COVID-19 pandemic closed operations of APOGEE-2S from March 2020 to October 2020 and after reopening, the APOGEE-2S observations spanned an ideal time of the year to focus on the Magellanic Clouds. 
A total of 23 new fields were added, placing five 12-visit fields in the SMC (named SMC$8-12$) and the remaining 18, 9-visit fields placed throughout the LMC (named LMC$18-35$).
This increases the SMC coverage by 70\% and the LMC coverage by 106\%.
The target selection for these fields are identical to that \deleted{described in Zasowski et al. (2017) and Nidever et al. (2020) and summarized above.}
\added{summarized above \citep[see also][\citetalias{zasowski2017}]{nidever2019a}.}

\noindent \textbf{\textit{Optical Calibration:}}~ 
The APOGEE-2S Magellanic Cloud program is one of the largest spectroscopic surveys of the Magellanic System, but there was very little overlap with prior studies in the optical to which the APOGEE \added{near-infrared-based} measurements could be directly compared.
To rectify this, we added an APOGEE field that overlaps with the study of \citet{vanDerSwaelmen2013} that measured stellar parameters and \replaced{chemical}{elemental} abundances from optical spectroscopy. 
While the APOGEE-2S targeting primarily focused on the most luminous stars in the LMC, \citeauthor{vanDerSwaelmen2013} observed stars lower on the giant branch. 
The purpose of this field is to characterize any differences between the \deleted{targeting and pipelines between} \added{target selection of} APOGEE-2 and \citet{vanDerSwaelmen2013}\added{(e.g., upper versus mid-giant branch and results derived from infrared spectra).
Some early evaluations of these differences are given in \citet{nidever2019a}.}

This field is named `LMC\_VdS' with a field center at $(\ell,b)=(283,-34)$ and was observed using a single cohort in a 24-visit design. 
Using a plate radius of 1.0\degs, 67 targets from \citet{vanDerSwaelmen2013} were prioritised to \added{a limiting magnitude of} $H=14.5$\,mag, with three stars rejected due to fiber collisions.
The targets from \citet{vanDerSwaelmen2013} will have targeting bit 22 set in \texttt{APOGEE2\_TARGET1} because they are confirmed members.
The remaining fibers in this field were selected following the \added{standard} LMC target selection \deleted{Nidever et al. (2020) that is summarized above}.
The corresponding depth for both of these targeting schemas is $H$=14.5 and this is fainter than the magnitude limit needed to reach our nominal $S/N$=100 goal in a 24-visit field (see \autoref{tab:maglims}). 

\noindent \textbf{\textit{LMC Substructure:}}~
A Contributed Program (see \autoref{sec:monachesi}) targeted the LMC substructure and these observations were coordinated with the Magellanic Clouds Working Group. 
As part of the COVID-19 extension, these were folded into Main Survey and each plate was allocated additional visits to reach the survey $S/N$ goal at 9-visits. 
No changes were made to the targeting bits or \programname\ to reflect this change.
For details on this program see \autoref{sec:contributedprograms} for the general implementation of Contributed Programs and \autoref{sec:monachesi} specifically.

\added{In all survey fields focused on the Magellanic Clouds, the following targeting bits are used. 
Confirmed Magellanic Cloud members have targeting bit 22 set in \texttt{APOGEE2\_TARGET1}.
Potential Magellanic Cloud stars, with candidates typically identified using photometric criteria, have targeting bit 23 set in \texttt{APOGEE2\_TARGET1}.
The \texttt{PROGRAMNAME} for this program is ``magclouds.'' }

\subsection{Open Clusters} \label{sec:openclusters}

Open Clusters are important tools to study the chemical patterns and evolution of the disk and targeting such objects has been a major component of APOGEE-1 and APOGEE-2. 
In particular, the Open Cluster Chemical Abundance and Mapping (OCCAM) Survey has worked to ensure that as many open clusters as possible are included in the APOGEE footprint \citep[e.g., ][]{frinchaboy2013,donor2018, donor2020}.
In DR16, \citet{donor2020} found 128 clusters had some observations with 71 clusters having sufficiently many members to be used to study chemical gradients in the disk. 

Targeting for the Open Clusters proceeds in a series of steps that are explained in detail in the OCCAM papers \citep[][]{frinchaboy2013,donor2018, donor2020} as well as \citetalias{zasowski2013} and \citetalias{zasowski2017}.
First the cluster mean proper motion is determined using a kernel convolution method.
Stars closer to the cluster average proper motion value are given higher priority and, where available, spectroscopic and/or kinematic member stars from previous studies are targeted. 
Finally, photometry is used to remove clear non-members from the color-magnitude diagram.
Stars selected as potential open cluster members have the targeting bit 9 in \texttt{APOGEE2\_TARGET1} set.

\noindent \textbf{\textit{Additions to the Open Cluster Sample:}}~
The goal in supplementing the Open Cluster sample was specifically to target distant clusters in both the inner- and outer- Galaxy to improve measurements of the Galactic abundance gradient \citep[e.g., ][]{frinchaboy2013,donor2018, donor2020}.
Additionally, clusters were included to help expand our cluster-based calibration efforts for both the metal-rich and \added{relatively} metal-poor extremes \added{that can be probed by Open Clusters} \citep[see e.g.,][]{holtzman_2018}. 
We added six open clusters amounting 20 visits; these clusters are Berkeley~75, Berkeley~81, NGC~5999, NGC~6583, NGC~6603, and Tombaugh~2.

\noindent \textbf{\textit{Main Sequence Calibrators in M\,67:}}~
The COVID-19 extension opened up some time during which the open cluster M\,67 was observable.  
APOGEE-2N designed a field dedicated to observing down the main sequence of M\,67 to build a calibration sample of M~dwarfs (see \beatonetalpar).  
However, this 36-visit field could not be completed in the COVID19-modified schedule for APOGEE-2N\added{, and only 10 visits were performed in this field.}  
Because this important calibration field could not be finished in APOGEE-2N, and time was available with the COVID-19 extension to APOGEE-2S, the decision was made to re-design this northern field for southern observation. 
While this required changing the targets slightly from the original design, which is described in \beatonetal\, due to the difference in plate scale \added{and field-of-view} between APOGEE-2N and -2S, the target selection and design follows the description given there \added{though we note that the targets are not one-to-one identical}.

\subsection{\ktwo} \label{sec:k2}

APOGEE-2N initiated a large-scale campaign to obtain complementary APOGEE spectra in the \ktwo\ campaigns (described in \beatonetalpar). 
The full program was difficult to accommodate fully in APOGEE-2N and, with the goal of providing the most complete possible scientific dataset leveraging \ktwo, we transferred 87 1-visit fields from six different \ktwo\ campaigns to the APOGEE-2S field plan. 
At the time of this decision, comparisons of the relative performance of the ASPCAP pipeline on targets observed both with APOGEE-2N and APOGEE-2S suggested that splitting the program would not cause bias in the \ktwo\ scientific results \citep[a full discussion is given in][]{jonsson_2020}. 
The fields transferred to APOGEE-2S plan correspond to Campaigns C8, C10, C14, C15, C17, and half of the fields from C6. 
We note that the differences between the field-of-view and plate-scale mean that the plates for APOGEE-2N and APOGEE-2S are not identical \added{, and this should be taken into account if data from both telescopes is combined}. 

We briefly summarize the targeting here, noting that a more involved discussion is given in \beatonetal.
The targeting made use of knowing what stars were already targeted by HERMES \citep{Sheinis_2015,Sheinis_2016} using the setup used in the GALAH survey \citep{Wittenmyer_2018,Sharma_2019}. \deleted{and oscillator identification following Hon et al. (2019)} \added{In addition, asteroseismic oscillator identification following the data processing and analysis of \citet{Hon_2019}, but applied to \ktwo\ data was performed before targeting; these asteroseismic results have been published in \citet{Zinn_2020} for three \ktwo\ campaigns}.
From all stars with \ktwo\ data, we first remove stars with existing APOGEE observations (APOGEE-1 or APOGEE-2).  
From here the targets were prioritized as follows: 
\begin{enumerate} \itemsep -2pt
    \item known planet hosts (\texttt{APOGEE2\_TARGET2} bit 11),
    \item stars with a confirmed oscillation or granulation signal (\texttt{APOGEE2\_TARGET1} bit 30),
    \item red giants targeted by the \ktwo\ Galactic Archaeology Program \citep[GAP;][\texttt{APOGEE2\_TARGET2} bit 0]{Stello_2017} and \textit{not} observed with HERMES, 
    \item GAP targets observed by HERMES \citep[see][\texttt{APOGEE2\_TARGET2} bit 17]{Wittenmyer_2018,Sharma_2019}, 
    \item M-dwarfs in the unbiased sample from the APOGEE-2N ancillary program (\texttt{APOGEE2\_TARGET3} bit 28).
    \item stars meeting the criteria for the ``main red star'' sample (\texttt{APOGEE2\_TARGET1} bit 14). 
\end{enumerate}
While the target selection was prioritised following this scheme, an individual target could be present in multiple target categories.
Field centers were optimized to reach the most targets weighted by priority. 

All \added{APOGEE-2N} \ktwo\ observations will have the \texttt{PROGRAMNAME} `k2\_btx'
and field names like `K2\_C\#\_$lll$$\pm$$bb$\_btx', where C\# indicates the \ktwo\ campaign and $lll$$\pm$$bb$ indicate the Galactic coordinates for the field center.
\added{For APOGEE-2S \ktwo\ observations the \programname\ is `k2' and the field names have format `K2\_C\#\_$lll$$\pm$$bb$.'}
All targets in \ktwo\ fields have \texttt{APOGEE2\_TARGET3} bit 6 set, as well as specific flags for the sub-targeting category as detailed above.

\subsection{Globular Clusters} \label{sec:globularclusters}
The original targeting of globular clusters in APOGEE-2S is described generally in \citetalias{zasowski2017} \deleted{with more details given in Meszaros et al. 2010}.
While there were no changes to the globular cluster program during the main survey of APOGEE-2S, the COVID-19 extension allowed for the expansion of observations on four globular clusters: NGC\,1851, NGC\,2808, M\,79, and $\omega$ Centauri ($\omega$Cen), and the design of a new globular cluster field, NGC\,2298, in total summing to 30 new visits, with six per cluster. 
For NGC\,1851, NGC\,2808, and M\,79, new designs were created to target known members from \citet{sollima2020} that are brighter than $H = 12.8$\,mag.  
A new field (\texttt{FIELD} of N2298) was made to target the globular cluster NGC\,2298, and similarly, targets for this cluster were taken from \citet{sollima2020} with the same magnitude limit.  
The member stars from these \replaced{four}{five} clusters are flagged with \texttt{APOGEE2\_TARGET2} bit 10.

\subsubsection{Omega Centauri}
In the original targeting of $\omega$Cen, the acquisition camera, with its 5.5$'$ occlusion radius, was placed near the center of the cluster, blocking the core of the cluster from observation.  
This notably leaves a significant gap in the APOGEE-2 spatial coverage of $\omega$Cen\added{, with the central 5.5$'$ of Omega Centauri containing no APOGEE targets due to the central obscuration (see \autoref{fig:apo2s_plate})}.  
In addition, since the design of the original $\omega$Cen field, \citet{ibata2019} used \gaia~DR2 \citep{gaia-dr2} observations to reveal a thin tidal stream coming off of $\omega$Cen.  
This too motivated observing more of $\omega$Cen.

With some of the time made available in the COVID-19 extension of APOGEE-2S, it was decided to allocate some of the available time to better cover $\omega$Cen and fill in this \added{spatial coverage} gap (by moving the field center, and therefore the location of the acquisition camera to a different position on the sky), albeit not to the same depth as originally designed.  
Instead of the 24 visits assigned to the original $\omega$Cen field --- which were split into two sets of long cohorts that received 12 visits each, with short and medium cohorts receiving 3 and 6 visits, respectively --- the new $\omega$Cen field (\texttt{FIELD} named `Omegacen2') was designed to have 6 visits, which were split into two sets of medium cohorts that received 3 visits each, with short cohorts receiving 1 visit each.  The minimum magnitude for this field was lowered to $H = 12.8$ for special targets, instead of the nominal $H = 12.2$ for a 3-visit faint cohort, \deleted{because as with some other updated programs,} $S/N$ $\sim 70$ was deemed sufficient.

The target selection for this new field, first, prioritized observing stars with high resolution optical spectra from \citet{johnson2020}, followed by proper motion identified $\omega$Cen members from \citet{vanleeuwen2000}, and finally $\omega$Cen stream candidates were prioritized last, however, because of their low sky density and distance from the center of the cluster all of the identified candidates were included. \deleted{Unfortunately, the identification of the $\omega$Cen by Ibata et al. (2019), was performed with dwarf stars, which were too faint for APOGEE-2S to observe at the distance of $\omega$Cen. So, }
Stream candidate giants were identified by consolidating stream candidates from \citet{anguiano2015}, \citet{fernandeztrincado2015}, and stars observed by the STREGA survey outside of $\omega$Cen's tidal radius and along $\omega$Cen's sequence in the $g_{\rm 0}$ vs $(g-i)_{\rm 0}$ color-magnitude diagram \citep[STRucture and Evolution of the GAlaxy with the VST;][]{marconi2014}.  
These latter stars were also required to have \gaia~DR2 proper motions within 1.6~mas yr$^{-1}$ of the bulk proper motion of $\omega$Cen $(\mu_{\alpha} \cos(\delta), \mu_{\delta}) = (3.24, -6.73)$ from \citet{baumgardt2019}. 
$\omega$Cen members with optical spectra or identified from their proper motions have been flagged with \texttt{APOGEE2\_TARGET2} bit 10 and the $\omega$Cen stream candidates have been flagged with \texttt{APOGEE2\_TARGET1} bit 19.


\begin{table*}
    \footnotesize
    \movetabledown=1.2in
    \centering
    \begin{rotatetable}
    \caption{Summary of APOGEE-2S Contributed Programs}
    \label{tab:external}
    \begin{tabular}{l c c r r r r c c l }
         Program &                                TAC & Program & Total &  Total &    Stars &  Average & Depth     & Subsection & Contact \\
         Name    &                                    &  Name   & Visits &  Stars & Observed &      SNR & $H$ [mag] &            &          \\
         \hline \hline 
         Star Forming Region G305             & CNTAC & borissova\_17a &     8  &    1000  &     250   &     71.4    &      12.2  &   \ref{sec:borisova}       & J.~Borissova \\ 
         Cool dwarfs in K2 fields             & CIS   & teske\_17a           &    43  &   10750  &    7250   &    134.9    &      11.0  &   \ref{sec:teske}          & J.~Van Saders \\
                                               &      & TeskeVanSaders\_17b &        &          &           &             &            &                            &  \\
         Upper Scorpius cluster               & CIS   & weinberg\_17a  &    98  &   21750  &    5106   &    143.0    &      11.7  &   \ref{sec:weinberger}     & A.~Weinberger \\
                                               &      & weinberg2\_17a &        &          &           &             &            &                            & \\
         Inner bulge and disk                 & CNTAC & zoccali\_17a &    17  &    3750  &    1750   &     91.4    &      11.7  &   \ref{sec:zoccali}        & M.~Zoccali  \\ 
                                               &      & zoccali\_18b &        &          &           &             &            &                            & \\
         Metal-poor stars in the inner Galaxy & CIS   & schlaufman\_17a &    18  &    4500  &    3000   &     97.1    &      11.0  &   \ref{sec:schlaufman}     & K.~Schlaufman  \\         
         The structure of the Ancient MW      & CIS   & kollmeier\_17a &    43  &   10750  &   10250   &     47.5    &      11.0  &   \ref{sec:kollmeier1}     & J.~Kollmeier \\       
         Low metallicity Cepheids             & CIS   & beaton\_18a &    18  &    1500  &    1500   &    195.1    &      12.2  &   \ref{sec:beaton}         & R.~Beaton \\ 
         TESS/APOGEE survey                   & CIS   & TeskeVanSaders\_18a &   154  &   38500  &   26500   &    168.1    &      11.0  &   \ref{sec:teskevansaders} & J.~Teske,  \\
          \multicolumn{9}{c}{}                                                                                                                 &  J.~Van Saders \\ 
          \multicolumn{9}{c}{}                                                                                                                 &  and R.~Beaton \\      
         Carina Star Forming Complexes        & CNTAC & RomanLopes\_18a &     5  &     750  &     750   &    175.3    &      12.2  &   \ref{sec:romanlopes}     & A.~Roman-Lopes \\
         Corona Australis protocluster        & CNTAC & stutz\_18a &     4  &     500  &     500   &    135.9    &      11.7  &   \ref{sec:stutz1}         & A.~Stutz \\ 
         Bulge globular clusters              & CNTAC & geisler\_18a, geisler\_19a &    57  &    5561  &    3217   &    106.9    &      12.8  &   \ref{sec:geisler}        & D.~Geisler \\
                                              &       & geisler\_18b, geisler\_20a &        &          &           &             &            &                            & \\
         Massive Stars in the SMC and LMC     & CIS   & Drout\_18b &    22  &    5250  &    3500   &    130.5    &      11.7  &   \ref{sec:drout}          & M.~Drout \\
         Rossete Molecular Clouds             & CNTAC & stutz\_18b &     4  &     370  &     370   &     56.5    &      12.2  &   \ref{sec:stutz2}         & A.~Stutz \\ 
         Carina Protocluster                  & CNTAC & stutz\_19a &     6  &     750  &     500   &     66.3    &      11.7  &   \ref{sec:stutz3}         & A.~Stutz \\ 
         OB Stars                             & CIS   & kollmeier\_19b &    42  &    2750  &     500   &    160.6    &      12.8  &   \ref{sec:kollmeier2}     & J.~Kollmeier \\
         \multicolumn{9}{c}{}                                                                                                                                  &  A.~Tkachenko \\       
         LMC substructures                    & CNTAC & monachesi\_19b &    21  &    1500  &    1250   &     55.7    &      12.4  &   \ref{sec:monachesi}      & A.~Monachesi \\      
         Orion Nebula Cluster                 & CNTAC & stutz\_20a &     7  &    1000  &       0   &      0.0    &      11.7  &   \ref{sec:stutz4}         & A.~Stutz \\ 
         Carina Nebula Cluster                & CNTAC & medina\_20a &     7  &     750  &       0   &      0.0    &      12.2  &   \ref{sec:medina}         & N.~Medina \\ 
         NGC6362 cluster                      & CNTAC & fernandez\_20a &     2  &     500  &       0   &      0.0    &      11.0  &   \ref{sec:fernandez}      & J.~Fern\'andez-Trincado \\ 
         \hline \hline
    \end{tabular}
    \end{rotatetable}
\end{table*}

\begin{figure*} 
    \centering
    \includegraphics[width=\textwidth]{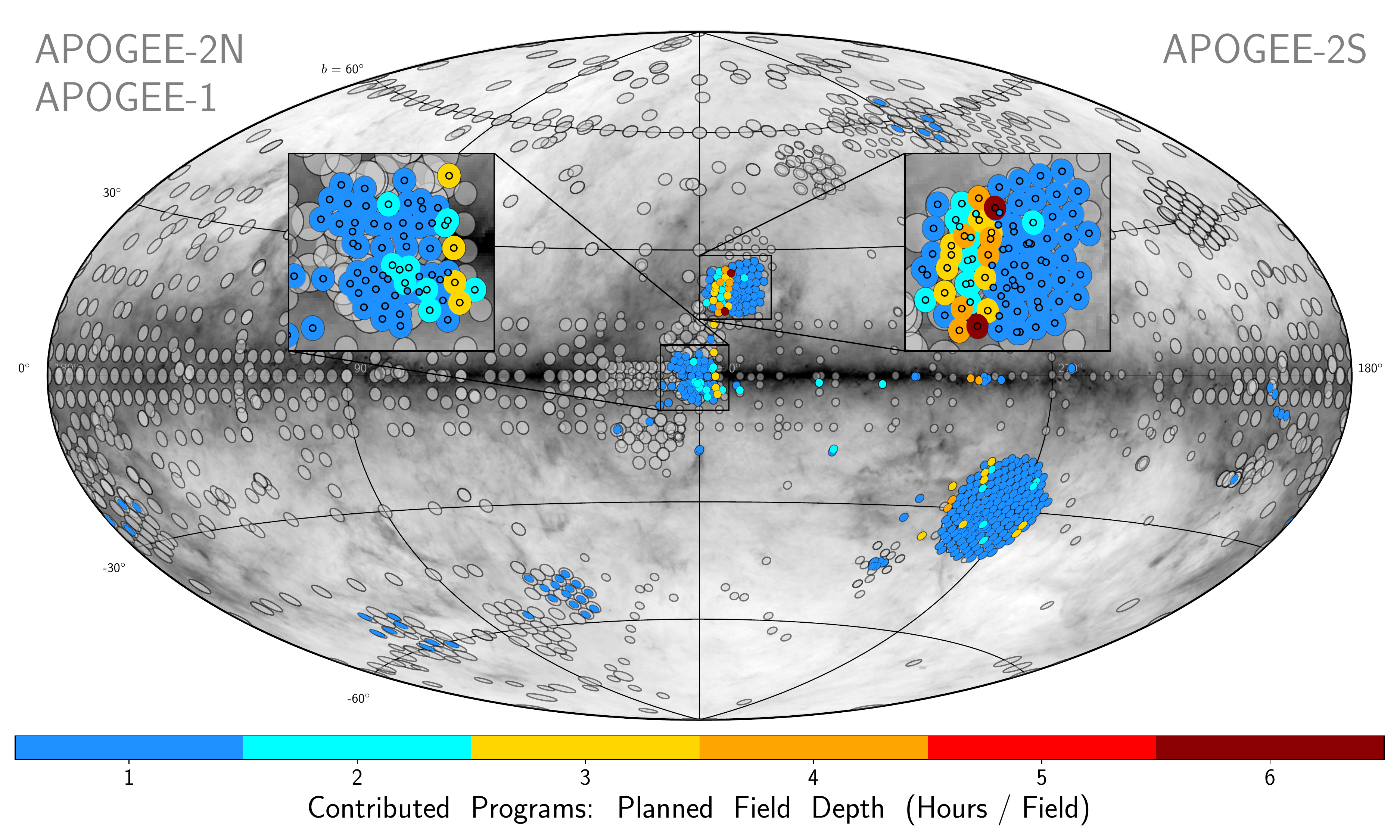}
    \caption{\explain{ This figure and caption have been updated}
    Sky distribution of APOGEE fields highlighting in color those from APOGEE-2S Contributed Programs. 
    The color-coding indicates the number of visits observed per field; grey symbols represent all other fields for context. 
    The background image is the dust map from \citet{schlegel1998}. 
    The inlays indicate the dense targeting in the Galactic center (left) and the Upper Scorpius star forming region (right), where we have overlaid circles to indicate the unique plate centers. 
    These fields represent $\sim$23\% of the APOGEE-2S observing program.}
    \label{fig:contributed_distribution}
\end{figure*} 

\section{Contributed Programs} 
\label{sec:contributedprograms}

Roughly 23\% of the observing time of the APOGEE-S spectrograph was used for programs that are independent from the main survey observing plan\added{and thereby do not follow the main survey targeting strategies}.
Such programs are allocated either through the Chilean Time Allocation Committee (CNTAC) or by the Observatories of the Carnegie Institution of Science TAC (CIS) and are both scheduled ``classically'' for specific night(s)\added{and observed ``classically'' (in the sense that time lost on a given night is only recovered through a subsequent TAC allocation).
From 2016 to 2020, 104 nights were used for Contributed Programs following the ratio of Main Survey versus Contributed Program nights of approximately 3:1 per annum}.

The targeting scheme of each of these programs is entirely decided by its Principal Investigator (PI) \added{via the submission of special targets; this includes what is observed, for how long and, if applicable, at what cadence}. 
The APOGEE-2S team only applies some basic restrictions to the target lists to ensure that the plates can be plugged and are observable during the night(s) assigned to the program.

The PI(s) of each project may choose whether the observed data would be included in the SDSS Data Releases, known as ``Contributed Programs,'' or if the data would remain external to the survey, known as ``External Programs.''
For Contributed Programs, while the data are collected according to the specifications of the TAC and the PIs, the  spectroscopic data, itself, is treated identically to that of the main survey. 
More specifically, the data are processed with the existing APOGEE data processing pipelines, are made into the same data products, and are made available to the collaboration in the same way.
\added{For the duration of APOGEE-2S,} all of the PIs chose for their observations to be executed as ``Contributed Programs.''

Table \ref{tab:external} provides the main observational aspects of the 19 Contributed programs, including the nominal depth calculated based on the maximum visits received per star. 
\autoref{fig:contributed_distribution} provides the sky distribution of Contributed Programs.
Each field is color-coded by the planned field depth (recall, 1~visit is roughly 1~hour of on-sky integration) to provide a sense of the targeting strategy. 
The two insets in \autoref{fig:contributed_distribution} zoom in on particularly densely targeted regions of the sky.

Field naming conventions from the main survey were adopted for these programs, but the fields from Contributed Programs have ``-C'' and ``-O''appended to their name, for CNTAC and CIS programs, respectively.
Fields from the \tess\ CIS program specifically (TeskeVanSaders\_18a; \autoref{sec:teskevansaders}) have a further ``\_TESS'' appended at the end of their names.
Targets included in these observations have bit 24 and 25 set in \texttt{APOGEE2\_TARGET2} flag for CIS and CNTAC programs, respectively.

Because the targets are not selected following standard APOGEE-2 procedures, the programs are \replaced{identifiable.}{not identifiable using targeting bitmasks.} 
Each program is identified by its ``program name'' in the \programname\ tag (given in \autoref{tab:external}). 
The program name is typically constructed as the last name of the PI followed by the first semester it was allocated; because of the classical-style implementation of these programs, many programs spanned multiple observing semesters.

\added{In the sections that follow, we provide descriptions of the 19 individual Contributed Programs in regards to its observations and science goals, contact scientists, and the awarding time allocation committee.
Just as for the survey descriptions, these correspond to the program as designed and do not reflect the level of completeness obtained by the program either by number of fields, number of stars, or intended $S/N$.}

\subsection{Star Forming Region G305} \label{sec:borisova}

Contact Information: J.~Borissova (Universidad de Valpara\'iso). \\ 
Located within the Scutum-Crux arm of the Milky Way ($\ell$=305.4\degs, $b$=0.1\degs), with projected diameter of $30\,\rm{pc}$, the G305 star-forming complex is both relatively nearby ($\sim3.8\,\rm{kpc}$) and one of the most luminous H{\sc II} regions in the Milky Way \citep{urquhart2014}. 
The region contains several distinct sites and epochs of star formation, which permits the study of massive star formation, massive star evolution, and the impact of these processes on the surrounding environment. The morphology of the region and positions of star formation indicators suggests that interaction between the evolved massive stars and remnant, natal molecular material is taking place, resulting in ongoing star formation activity. Thus, G305 region is a perfect test-bed object, both for intermediate and high mass star formation. 

The scientific aim of this project is to analyze the kinematics and metallicity of stars in G305 region. 
The project will combine the homogeneous, high precision photometry and variability from VVV \citep{minniti2010}, astrometry from \gaia~DR2 \citep{gaia-dr2}, and radial velocity and metallicity measurements from APOGEE-2. 
With this data, the project will:
    (1) improve the census of both massive and intermediate-mass stars in the area; 
    (2) measure the kinematics of different stellar groups (YSOs, OB, WR stars) and confirm/reject memberships of the stars to the clusters projected in the region; 
    (3) outline the potential spatial substructures and measure the star formation histories to probe models of triggered versus passive star formation; and 
    (4) calculate the present-day star formation rate in the area using the method of direct counts of YSOs as compared with other star forming regions in the Milky Way. 
Some results from these observations are published in \citet{borissova2019}.
Observations from this project have \programname\  borissova\_17a.

\subsection{Cool Dwarfs in \ktwo\ fields} 
\label{sec:teske}

Contact Information: J.~van Saders (University of Hawaii) and J.~Teske (Carnegie Institution For Science, Earth and Planets Laboratory). \\ 
The goal of this program is to collect APOGEE-2S spectra of bright, cool dwarf stars that are being monitored by the \ktwo\ ecliptic survey \citep{howell2014}, including stars known to host planet candidates through the analysis of their light curves.
These data will be used to measure stellar parameters and chemical abundances for low-mass stars in the \ktwo\ fields.
The scientific goals are: (1) to determine rotation-based ages, (2) to constrain the age-metallicity relation, and (3) look for a correlation between exoplanet architecture and host-star composition.
In particular, M~dwarf spectra are of high value in the development of an analysis method for extracting reliable, detailed abundances of $T_{\rm eff} < 4000\,K$ stars.
This project aims at observing 34 fields from six \ktwo\ Campaigns: C1, C2, C3, C4, C7, and C8.
Observations from this project have \programname\ teske\_17a, or TeskeVanSaders\_17b.

\subsection{Upper Scorpius Cluster} 
\label{sec:weinberger}

Contact Information: A.~Weinberger (Carnegie Institution For Science, Earth and Planets Laboratory). \\
This program aims to survey the young ($\sim 10\,{\rm Myr}$ old), Upper Scorpius cluster to measure the properties of a large, homogeneously targeted population of young stars.
We primarily target cluster members that were included in  \ktwo\ and for each star we intend to measure \teff, \logg, and radial velocity.

The targets for this project come from three separate sources:
\begin{enumerate}
    \item we included all targets fainter than $H$=7.5 from three \ktwo\ guest observer programs during Campaign 2\footnote{\url{https://keplerscience.arc.nasa.gov/k2-approved-programs.html\#campaign-2}} that focused on Upper Sco. 
    These are 
    Program 2052 (PI Covey: \textit{\ktwo\ Monitoring of Confirmed Members of Upper Sco and Rho Oph}), 
    Program 2057 (PI Hillenbrand: \textit{A \ktwo\ Study of Young Stars in Upper Scorpius}), and 
    Program 2063 (PI Kraus: \textit{Planets Around Likely Young Stars in Upper Sco and Rho Oph}). 
    These account for 1309 targets.  
    \item we included all additional (16) Upper Sco targets that were part of a ground-based parallax survey \citep{Donaldson_2017}; these were all fainter than $H$=7.5. 
    \item we included additional Upper Sco targets that were part of the \spitzer\ surveys of Upper Sco looking for circumstellar disks and were fainter than $H$=7.5 (78 targets) \citep{Carpenter_2005sptz,Carpenter_2006sptz,Carpenter_2006,Carpenter_2009}.
\end{enumerate}
Thus there were 1403 sources included in our master target list.  
When constructing input lists for the APOGEE-S targeting software, we allowed targets in the vignetted (0.8 - 0.95\degs radii) region of the field.

Observations from this project have \programname\ of weinberg\_17a, or weinberg2\_17a.

\subsection{Inner Bulge and Disk} 
\label{sec:zoccali}

Contact Information: M.~Zoccali (Pontificia Universidad Cat\'olica-Instrituto Milenio de Astrof\'isica). \\
The region called the ``Nuclear Bulge'' stands out from the main bulge due to several observed properties:
    (1) a much higher stellar density, with a clear break \citep{launhardt2002},
    (2) the presence of a dense interstellar medium \citep[the Central MolecularZone][]{morris1996}, 
    (3) ongoing star formation in scattered dense clouds, and 
    (4) a stellar Nuclear disk.
The the presence of a stellar nuclear bulge has been confirmed with APOGEE data \citep{schoenrich2015}, however many open questions remain regarding its formation and relationship to other bulge structures.

To sample this region with APOGEE, five fields were designed at the following Galactic coordinates ($\ell$,b):  (-4,0), (-2,0), (0,0), (2,0), and (4,0), respectively.
Targets were selected from imaging from VISTA Variables in the Via Lactea \citep[VVV;][]{minniti2010}, but with PSF photometry derived by Rodrigo Contreras (priv.~communication) with the method described in \citet{contreras-ramos2017}.
Because the interstellar extinction towards the nuclear bulge is extremely large ($A_{\rm V}>20$) the observed red giant branch has a very broad distribution in color.
To define a target box in color-magnitude space that creates a uniform sampling of the stellar populations at different metallicities, an initial selection of stars with 2.4$<E(J-K_{\rm s})<$3 was imposed, using the reddening maps from \citet{gonzalez2018}. 
In the central field, this color-cut excluded stars within $\pm$0.3\degs of the Galactic plane, but permitted the CMD box with $H<11.8$ to span the full RGB color-range for the other four fields without imposing a significant color bias.
If not designed-out of the sample selection, a color bias could introduce a metallicity bias; we note that the color-range is different for each of the fields.

The APOGEE observations will be used to derive the metallicity distribution function for these regions and it will be used to test if the Nuclear disk is the central extension of the inner bulge, or if is a distinct stellar component.
Furthermore, an empirical metallicity distribution function will break the degeneracy in the derivation of the star formation history of this region, by comparing the available deep NIR CMDs \citep{nogueras-lara2020} with synthetic color magnitude diagrams.
Observations from this project have \programname\ of zoccali\_17a, or zoccali\_18b.

\subsection{Metal-poor Stars in the Inner Galaxy} 
\label{sec:schlaufman}

Contact Information: K.~Schlaufman (Johns Hopkins University). \\
The oldest stars in the Milky Way are very metal-poor (i.e., $[\mathrm{Fe/H}] < -2.0$) and may be found in the inner Galaxy \citep[e.g.,][]{tumlinson2010}.  
However, extreme reddening, extinction, and crowding in the inner Galactic bulge have made finding metal-poor stars in this region impossible with traditional techniques.  
\citet{schlaufman2014} published an efficient metal-poor star selection that uses only infrared photometry, thereby overcoming the previous barriers to the identification of metal-poor stars in the inner Galaxy.  
The targets in this program were selected following the pure-infrared metal-poor star selection from \citet{schlaufman2014} using the photometry from the \emph{Spitzer}/IRAC Galactic Legacy Infrared Mid-Plane Survey Extraordinaire \citep[GLIMPSE;][]{benjamin2003,churchwell2009} and applying dereddening from high-resolution bulge-specific maps from \citet{gonzalez2011,gonzalez2012} assuming the \citet{nishiyama2009} extinction law.  
An restriction of $H \leq 12.5$ was applied. 
Observations from these project have \programname\ schlaufman\_17a.

\subsection{The Structure of the Ancient Milky Way}
\label{sec:kollmeier1}
Contact Information: J.~A.~Kollmeier (Observatories of the Carnegie Institution for Science) and R.~Poleski (Warsaw). \\ 
RR Lyrae (RRL) stars provide a unique view into the ancient interior of the Milky Way.  
In addition to being old stars, the pulsations of RRL stars have long allowed accurate distance estimates. 
The distance estimates can be combined with proper motions \citep{Poleski_2013,Poleski_2015} and radial velocities (from APOGEE) to obtain full six dimensional phase-space information for these stars.  
The goal of this project is to target a large sample of RRL stars identified by the Optical Gravitational Lensing Experiment (OGLE) survey in Galactic bulge \citep{udalski1992} for radial velocity measurement with APOGEE-2S. 
In particular, this program was designed to penetrate the dust of the Milky Way to see the 3-D structure of it's old central component -- to compare with theories of formation of the bulge.
Targets were selected by applying the OGLE-defined periods to the $H$-band period luminosity relationship to predict the mean $H$-band magnitudes. 
All the RRL stars from this program have bit 24 set in \texttt{APOGEE2\_TARGET1}.
Observations from this project have \programname\ kollmeier\_17a.

The original main survey plan also included RRL observations \citepalias{zasowski2017}, designed by the same Contributed Program PIs.  
Thus, for the purposes of analysis and selection, this Contributed Program and the main survey program are identical.

\subsection{Low Metallicity Cepheids} 
\label{sec:beaton}

Contact Information: R.~Beaton (Princeton University and the Observatories of the Carnegie Institution for Science). \\ 
This Contributed Program has collected single-phase APOGEE-2S spectroscopy for the brightest and most well-studied Cepheids in the LMC.
The APOGEE measurements can be combined with literature multi-band optical to mid-infrared photometry to study trends metallicitiy and the absolute magnitude dispersion of these stars against the mean Leavitt Law. 
The Cepheid sample was largely selected from the OGLE-III catalog of Cepheid variables in the LMC \citep{Soszynski_2008}, but was supplemented with key calibrators used for the Cepheid-based extragalactic distance scale \citep{Persson_2004,Scowcroft_2011,Scowcroft_2016}. 
To obtain high $S/N$ data, these stars are observed in two to three sequential visits and, thus, pipeline combined data is all at the same phase of the photometric light curve. 
As a result, the targets go to an $H$ depth of a 3-visit field (using the mean magnitudes; \autoref{tab:maglims}).
Additional fainter Cepheids were also targeted as part of this program, with the caveat that they may not reach $S/N$ of 100 per pixel.
We caution that sometimes the observations were significantly distributed in time.
The remaining fibers were back-filled with targets from the ``TESS/APOGEE Survey'' (\autoref{sec:teskevansaders}) but only going to a depth of $H\sim11$\,mag.
Observations from this project have \programname\ beaton\_18a.

\subsection{TESS/APOGEE Survey} \label{sec:teskevansaders}

Contact Information: J.~Teske (Carnegie Institution For Science, Earth and Planets Laboratory), J.~van Saders (University of Hawaii), and R.~Beaton (Princeton University and the Observatories of the Carnegie Institution for Science). \\
This program intended to collect APOGEE-2S spectra of bright ($7<H<11\,{\rm mag}$) stars that fall within the $450\,{\rm deg^2}$ of the Southern Ecliptic Pole, a region of the sky where NASA's \tess\ satellite \citep{ricker2015} has had continuous 1 year coverage, and with \tess\ extended mission will have an additional year of coverage. 
These APOGEE-2S data will be used to measure stellar parameters and chemical abundances.
When combined with the precise \tess\ photometric light curves, these data will enable investigations of open questions about exoplanets, stellar astrophysics, and distance ladder calibration. Target selection was made via simple 2MASS color and magnitude cuts: $7<H<11$ and $J-K > 0.3$.
As plates were drilled, target priorities were defined as: 
    (1) subgiants where asteroseismic detections are expected, 
    (2) asteroseismic dwarfs from the ATL \citep{scholfield2019}, 
    (3) targets likely to be placed on 2 minutes cadence for planet searches based on the CTL \citep{stassun2018}, 
    (4) hot star seismic targets (exceptions to the color cut), and 
    (5) bright giants.
Fields from this program have the word ``-O\_TESS'' appended at the end of their field name.
Observations from this project have \programname\ TeskeVanSaders\_18a.

\subsection{Carina Star Forming Complexes} \label{sec:romanlopes}

Contact Information: A.~Roman-Lopes (Universidad de La Serena). \\
The Carina star-forming complexes contain some of the most massive star-forming regions in the Milky Way.
It is an ideal laboratory to test theories of the spatially segregated formation of stars and triggering mechanisms.
This program takes advantage of the large FOVand multiplexing capability of the APOGEE-S spectrograph to confirm the massive nature of hot star candidates in the Galactic plane. 
The fields observed in this project are located in the vicinity and periphery of regions forming massive stars in the Carina star-forming complexes.
The excellent spectral resolution provided by the APOGEE-2S instrument, and the high $S/N$ radial velocity measurements enabled us to obtain spectroscopic information of about $1,750$ candidate massive stars, a task virtually impossible using single slit NIR spectroscopic facilities.
Further details for this program are provided in \citet[][]{roman-lopes2020}.
Observations from this project have \programname\ RomanLopes\_18a.

\subsection{Corona Australis Protocluster}
\label{sec:stutz1}

Contact Information: A.~Stutz (Universidad de Concepci\'on).\\
This program aims at measuring the radial velocities of the young stellar population in the Corona Australis protocluster.  
The "Slingshot" mechanism proposes that young stars gain kinetic energy via gas filament oscillations \citep{Stutz_2016,boekholt_2017, Stutz2018a}.  
Ultimately, such oscillations may ``eject'' protostars from their dense gas cradles, cutting off the supply of gas to the new stars.  
The ``Slingshot'' scenario proposes that magnetically driven instabilities in the gas may be responsible for the oscillations. 
Whatever the ultimate oscillation mechanism, the APOGEE-2S Corona Australis RV measurements enable the dynamics of this proto-cluster to be compared to those of the Orion-A Integral Shaped Filament.  

Targets for this program were selected from the \emph{Spitzer}-based catalog of Corona Australis members assembled by \citet{Peterson_2011}. 
We prioritize all young stellar objects (Classes 0, I, II, and III) with $H<$~14, comprising a high-priority sample of 50 objects that are distributed across two plates to cover the full stellar population.
Observations from this project have \programname\ stutz\_18a.

\subsection{CAPOS: the bulge Cluster APOgee Survey} \label{sec:geisler}

Contact Information: D.~Geisler (Universidad de Concepci\'on and Universidad de La Serena).\\ 
The Galactic bulge hosts a large number of globular clusters \citep[e.g.,][]{Harris_2010,Kharchenko_2016}.
These clusters are powerful cosmological probes to investigate the formation and chemical evolution of this key Galactic component \citep[APOGEE studies in individual stars include:][]{hasselquist_2020,rojas-arrigada_2020,Queiroz_2020}.
Unfortunately, until recently, we have not been able to unleash the full power of the bulge globular clusters to help unravel its mysteries due to their high extinction, which strongly impedes optical observations \citep[e.g.,][]{Gran_2019,Palma_2019}.
However, extinction effects are minimized by observing in the NIR, allowing us to fully exploit the bulge GC's extraordinary archaeological attributes.
The bulge Cluster APOgee Survey (CAPOS) will observe 20 of the bona-fide bulge globular clusters that were not included for the APOGEE-2 observations \citep[see Z17;][]{Masseron_2019,meszaros_2020}.  
Combining this Contributed Program with the APOGEE Main Survey bring the total number of bulge clusters targeted with APOGEE to 28, just over 50\% of the known bulge clusters.
CAPOS will supply almost 75\% of the observed objects and build a legacy database of the bulge globular cluster system. 
This will provide much better and more self-consistent radial velocities and spectroscopic metallicities than currently available for all the observed bulge globular clusters. 
These observations will permit investigation of many salient details of these systems like their chemistry, existence and properties of multiple populations, and orbital properties, among others. 
Observations from this project have \programname\ of geisler\_18a, geisler\_19a, geisler\_19b, or geisler\_20a.

\subsection{Massive Stars in the SMC and LMC}
\label{sec:drout}

Contact Information: M.~Drout (University of Toronto). \\ 
Our goal was to conduct a wide and shallow survey of the evolved massive star populations (e.g. yellow supergiants, red supergiants, and luminous blue variables) of the Large and Small Magellanic Clouds (LMC and SMC) with APOGEE-S.

For targeting, we followed a procedure similar to that outlined in \citet{Neugent_2010,Neugent_2012}, who study the red and yellow supergiants (RSGs and YSGs) of the LMC and SMC. We first selected all objects with 2MASS $H$ quality flag of `AAA' and magnitude of $H_{\rm 2MASS}$ \textless\ 12.2 that lie within 4\,deg$^2$ of the LMC and SMC, respectively. Centers for each galaxy were define as $(\alpha, \delta)_{\rm LMC}$  =  (05:23:34.5, -69:45:22) and $(\alpha, \delta)_{\rm SMC}$  = (00:52:44.8, -72:49:43).

To separate likely LMC/SMC supergiants from foreground dwarfs in the FOV we use a combination of \gaia-DR2 \citep{gaia-dr2} proper motions/parallax values and radial velocities/spectroscopic classifications, when available. We assign probabilities of membership based on \gaia-DR2 kinematics by the procedure described in \citet{ogrady_2020}, which is based on that outlined in \citet{helmi2018}. In brief, we use a curated sample of highly probable LMC/SMC members to define a distribution in ($\mu_\alpha, \mu_\delta, \pi$) followed by LMC/SMC members. Each 2MASS source described above is then cross-matched with \gaia~DR2, and a $\chi^2$ value is computed to assess the consistency of its proper motion and parallax with these distributions. For the purposes of targeting, we designate any star with a $\chi^2 < 6.25$ (indicating that the star falls within the region that contains 90\% of LMC/SMC members) as having \gaia\ kinematics that are ``consistent'' with LMC/SMC membership. In addition, we cross match all of the 2MASS sources with Simbad \citet{simbad_2000} in May 2018 in order to retrieve any existing radial velocity measurements for spectroscopic classifications. For the purposes of targeting we take any source with $v_{\rm rad}$ greater than 200\kms and 135\kms in the direction of the LMC and SMC, respectively, as having velocities consistent with membership in the Clouds \citep{Neugent_2010,Neugent_2012}, while we discard any source with a `V'-type classification as a foreground star.

With this ancillary data in hand, all APOGEE fields were then assigned targets, moving through an ordered list of priorities until fiber capacity was reached. The ranked priorities for the LMC were all follows. In the following definitions, we take the dividing line between 'blue/yellow' and ‘red’ stars to be $J-K$ = 0.9\,mag. This corresponds to an effective temperature of $\sim$4800\,K and is taken as the rough dividing line between YSGs and RSGs in \citep{Neugent_2012}.

\begin{itemize} \itemsep -2pt
    \item Star which are listed as a confirmed YSG or RSG in the spectroscopic sample of \citet{Neugent_2012} or as confirmed or candidate Luminous Blue Variables.
    \item Stars which are consistent with LMC membership based on both \gaia\ kinematics and an existing radial velocity, and have H \textless\ 11.3\,mag. 
    \item Stars which are consistent with LMC membership based on both \gaia\ kinematics and an existing radial velocity, and have $H$ \textless\ 11.8\,mag.
    \item Stars which are consistent with LMC membership based on \gaia\ kinematics, have no existing radial velocity data, and fall within the 2MASS color cuts defined by \citet{Neugent_2012} in order to identify YSGs down to a completeness limit of $\sim$12$M_{\odot}$, using the Geneva evolutionary models \citep{Maeder2001} and ATLAS9 atmosphere models \citep{Kurucz1993}. 
    \item Stars which are consistent with LMC membership based on \gaia\ kinematics, have no existing radial velocity data, and fall within the 2MASS color cuts defined by \citet{Neugent_2012} to identify RSGs with minimal contamination from AGB stars. 
    \item Other blue/yellow stars with $H$ \textless\ 11.6  that are consistent with LMC membership based on \gaia\ kinematics, have no existing radial velocity data. 
    \item Other red stars with $H$ \textless\ 10.8 that are consistent with LMC membership based on \gaia\ kinematics, have no existing radial velocity data.
    \item Any remaining stars with $H$ \textless\ 11.3 that are consistent with LMC membership based on \gaia\ kinematics, have no existing radial velocity data.
    \item Other blue/yellow stars with $H$ \textless\ 11.8 or red stars with $H$ \textless\ 11.6 that are consistent with LMC membership based on \gaia\ kinematics, have no existing radial velocity data.
    \item Other blue/yellow stars with $H$ \textless\ 11.6 or red stars with $H$ \textless\ 12.0 that are consistent with LMC membership based on \gaia\ kinematics, have no existing radial velocity data.
    \item Stars that are consistent with LMC membership based on \gaia\ kinematics and were classified as carbon stars in Simbad \citep{simbad_2000}. Note that objects with a listed spectral classification of `C' were explicitly excluded in all previous steps.
\end{itemize}

Priority listings for the SMC were entirely analogous. Main difference were than in Step 1 confirmed YSGs were added from \citet{Neugent_2010}, and in Steps 4 and 5 the $H$ magnitudes were shifted by 0.4\,mag in order to account for the difference in distance modulus between the LMC and SMC.

In total 16 LMC fields and 5 SMC fields were constructed and observed. 
Field centers were chosen to optimize the number of confirmed RSG, YSGs, and LBVs from the samples outlined in Steps 1 and 2, above, with over 97\% being successfully placed. 
When filling fields we allowed the stars from these samples to be placed in multiple fields, when possible, while all other stars were placed only once. 
In practice, this optimized the number of targets observed while also providing multiple epochs for a subset of stars over the $\sim$4~months in 2018-2019 when observations were carried out. 
Fields in dense regions of the clouds were typically filled proceeding through only the first 5 steps above, while some sparser regions proceeded though all 11 steps.

Observations from this project have \programname\ Drout\_18b.

\subsection{Rosette Molecular Clouds}
\label{sec:stutz2}

Contact Information: A.~Stutz (Universidad de Concepci\'on). \\
As with the Corona Australis program (see \autoref{sec:stutz1}), this program seeks to measure radial velocities for young stellar objects (YSO) to test the "Slingshot" model of star and cluster formation \citep{Stutz_2016,Stutz2018a,Stutz_2018b}.  
This program targets the protocluster associated with the Rosette molecular cloud.
Targets were selected from the WISE-selected catalog of cluster members assembled by \citet{Cambresy_2013}.
Cross-matching with \gaia~DR2 \citep{gaia-dr2} and 2MASS \citep{skrutskie2006}, targets were required to have parallaxes consistent with that of the complex (e.g., $d$ = 1.4--1.6 kpc) and prioritized according to their $H$ magnitude.
Sources with $H <12.2$ were assigned to a bright plate planned for a single 1-hour visit.
Members with 13.3$>H>$12.2, or suffering from a fiber collision with a brighter neighbor in the bright plate, were assigned to a `dim' plate planned for 3 hours of integration time.  
The faintest members (14.7$>H>$13.3) were assigned to spare fibers, but were expected to return spectra with very $S/N$.
Observations from this project have \programname\ stutz\_18b.

\subsection{Carina Protocluster}
\label{sec:stutz3}

Contact Information: A.~Stutz (Universidad de Concepci\'on). \\
These data will provide radial velocities (RV) for the young star protocluster population in the Carina cloud complex ($d\sim2.3\,{\rm kpc}$, $M_{\rm{gass}}\sim6.3 \times 10^{5}\, M_{\odot}$).
Carina's morphology and high-mass star content both stand in sharp contrast to the well studied Orion-A cluster.
One of the major differences is that Carina samples the complete initial mass function.
With $\sim90$ massive stars, Carina has a cosmological relevance as protocluster cloud at solar metallicity.
With high radial velocity precision, detailed scrutiny of the dynamics of the protocluster via self-consistent and simultaneous modeling of the stellar and gas kinematics is enabled.
APOGEE-2S provides the required radial velocity precision for this goal. 
By surveying the recently formed stars (${\rm ages}\sim2\, {\rm Myr}$) and combining with \gaia, the observational basis required for theoretical dynamical modeling of this representative protocluster-forming region is established.

Targets were selected from the \emph{Spitzer}-based catalog of Carina members assembled by \citet{Povich_2011}, after crossmatching to \gaia~DR2 \cite{gaia-dr2} and 2MASS \cite{skrutskie2006}. Sources with $H<$12 were prioritized for two-hours of integration, while fainter sources were planned for as many as 6-hours of integration time.
Observations from this project have \programname\ stutz\_19a.

\subsection{OB Stars}
\label{sec:kollmeier2}

Contact Information: J.~A.~Kollmeier (The Observatories of the Carnegie Institution for Science), A.~Tkachenko (KU Leuven), and C. Aerts (KU Leuven). \\
By obtaining high-resolution spectroscopy for intermediate- to high-mass stars (O and B spectral types), this program exploits the synergy between asteroseismology and stellar binarity to bring into focus key topics of stellar physics.
The ultimate goal is to obtain precise observational constraints on the angular momentum transport inside of these stars by combining asteroseismic inferences of the interior structure (where properties are determined through  the interpretation of their gravity-mode oscillations) with high-precision radial velocity measurements and atmospheric properties for these stars.
This sample was selected based on the type of variability imprinted on the high-precision \tess\ light-curves of the proposed targets, and the sample itself is a healthy mix between single and binary star systems.
The APOGEE-2S instrument plays critical role in this project by offering: 
    (i) preliminary orbital phase coverage for known binary stars, which will ultimately enable 
disentangling of individual spectral contributions and determination of atmospheric properties, and 
    (ii) another constraint on the binarity.
If stars are determined to be single, then the spectra will allow for determination of their atmospheric parameters.
Observations from this project have \programname\ kollmeier\_19b.

\subsection{LMC Substructures}
\label{sec:monachesi}

Contact Information: Antonela Monachesi (Universidad de La Serena). \\
The Magellanic Clouds are important systems to study.
Due to their proximity, we can investigate in great detail their formation, interaction, and evolution, which can be used together to place constraints on dwarf galaxy formation models.
During the last decade, an increasing number of photometric  surveys that cover both the main bodies of the Magellanic Clouds and their surroundings have uncovered many substructures extending out to several degrees from their centers \citep[e.g.,][]{mackey2016,mackey2018,pieres2017,nidever2019b}.
Such discoveries have especially revealed the complexities of the LMC.
More recently, \citet{belokurov2019} used \gaia\~DR2 \citep{gaia-dr2} to detect several substructures around the LMC out to a radius of 20$^{\circ}$.
These discoveries provide new insights that are fundamental to understand the formation and evolution of the Magellanic Clouds.
The spectral resolution and field-of-view of the APOGEE-2S instrument are crucial to analyze these newly discovered features and understand their origin.

To select these targets we used \gaia\~DR2 \citep{gaia-dr2} astrometry and 2MASS photometry \cite{skrutskie2006} available on these field targets.
The data was filtered using the following restrictions:
    (1) $\pi<0.2$~mas year$^{-1}$, 
    (2) fractional parallax uncertainty of $\sigma_{\pi}/\pi<3$, and 
    (3) a color-magnitude box of 0.6$<J-K_{\rm s}<$1.4\,mag, and 12$<H<$15\,mag.
Since this sample did not fill the plates, different proper motions cuts were used to select stars as back-fillers of lower priority.
Observations from this project have \programname\ monachesi\_19b.

\subsection{Orion Nebula Cluster \& Vela C}
\label{sec:stutz4}

Contact Information: A.~Stutz (Universidad de Concepci\'on). \\ 
This program consists of two parts, a program in Orion and in Vela~C.
The latter program was developed in addition to the former to use all of the time 
allocated to the program.
The Orion Nebula Cluster is forming within the Integral Shaped Filament  and serves as a laboratory for star and cluster formation physics \citep{Stutz_2018b}.
Based on APOGEE data, this team recently proposed the Slingshot model, in which the stars and cluster form on the oscillating Integral Shaped Filament \citep{Stutz_2016,Stutz2018a}.  
Even though previous APOGEE data have been pivotal to the development of this model, a large fraction of the young stars (470 stars brighter than $H=14$\,mag) do not yet have radial velocity measurements.
Moreover, a substantial fraction  of these are not detected in \gaia~DR2 \citep{gaia-dr2}.
The goal of this project is to obtain APOGEE-2S radial velocities to provide kinematic information for this young population.

We select young stellar objects (stars with disks) from the \emph{Spitzer}-based catalog assembled by \citet{Megeath2012}.  
We remove all young stellar objects that have been previously observed with APOGEE \citep[using both public and non-public data]{Cottaar_2014, DaRio_2017, Cottle_2018, Kounkel_2018}. 
We obtain a target list that is composed of sources
with $H > 13$\,mag, as the previous surveys adopted this as their limit.  
Because this region is highly clustered, we use multiple overlapping plates to build the sample due to fiber collisions.

To utilize the remainder of the assigned nights, additional plates targeting the Vela~C region were designed. These targets were selected using two sets of selection criteria: 
    (1) membership as assessed from \gaia~DR2 \citep{gaia-dr2} parallaxes and proper motions;
    (2) near- and mid-infrared spectral energy distributions, using photometry from 2MASS \citet{skrutskie2006} and \emph{Spitzer}, indicative of being on the pre-main sequence.
The SED-based selection was kindly provided by R.~Gutermuth (priv.~communication), following methods used to assemble the \emph{Spitzer} Extended Solar Neighborhood Archive \citep{Gutermuth_2019}.
Observations from this project have \programname\ stutz\_20a.

\subsection{Carina Nebula Cluster}
\label{sec:medina}

Contact Information: N.~Medina (Universidad de Valpara\'iso). \\ 
The stars in formation and early stages of evolution, so called young stellar objects, undergo rapid structural changes, such as formation and destruction of the disks, jets, envelopes, flares, accretion, etc., and for these reasons the variation of their luminosity and colours are widely observed.
Thus, the study of the photometric variability is one of the most powerful sources of astrophysical information. 
Variability in the infrared fluxes, which detects young stellar objects at very early, obscured stages of evolution, has been found to be a very common characteristic among young stars.
The ``VISTA Variables in the Via Lactea survey'' \citep[VVV;][]{minniti2010} and it's undergoing extension the``VISTA Variables in the Via Lactea Extended survey'' (VVVX) are time-domain infrared surveys which provide the unique opportunity to investigate the connection between infrared variability and its underlying physics.
Using the automated tool from \citet{medina2018} we searched the Carina region for variable young stellar objects in $K_{\rm s}$ and created a catalogue of candidate targets.
APOGEE-2 spectroscopy complements photometric variability studies.
The project will collect spectroscopy of $\sim$1000 young stellar objects, covering the whole projected area of the Carina Nebula Complex.
Spectroscopy allows validation of the photometric classification of the variable stars as young stellar objects, removing the AGB stars, novae and long period variables, which show similar light curve behavior, especially for high amplitude variables. 
Furthermore, the new  class of ``low amplitude eruptive variables'' proposed by \citep{medina2018} will be spectroscopically characterized for the first time.
Emission features in the APOGEE spectra may reveal a high accretion rate and, thereby, improve the classification obtained from the light curves. 
This can be used constrain the possible triggered physical process, as well as will investigate the kinematic distribution of young stellar objects in the Carina region.
The metallicity distribution of young stellar objects will allows for an outline of some possible differences in star formation process within the complex. 
The observations from this project have \programname\ medina\_20a.

\subsection{NGC\,6362 cluster}
\label{sec:fernandez}

Contact Information: Jos\'e G. Fern\'andez-Trincado (Universidad de Atacama). \\
NGC\,6362 is a nearby low-mass, old stellar system with an intermediate metallicity; as such, it is a perfect tool to study the phenomenon of multiple stellar populations to large projected radius. 
Measurements from APOGEE-2S will place constraints on the extra-tidal population and, thus, will provide insight into its formation, evolution, chemistry, and kinematics, all of which are poorly understood at present.
There was no compelling evidence for any significant extra-tidal population in NGC\,6362 until recent work by \citet{kundu2019}.
\citet{kundu2019} claimed to have identified extra-tidal candidates using photometry and astrometry from \gaia\~DR2 \citep{gaia-dr2} using color-magnitude and proper motion constraints that maximized the contrast between cluster and field populations. 
This program aims to observe the bulk of the \gaia\~DR2 extra-tidal candidates as well as supplementing the current APOGEE-2S sample of potential cluster members to trace the chemical and dynamical history of NGC\,6362
Observations from this project have \programname\ fernandez\_20a.

\section{Summary}
\label{sec:summary}

The second generation of the APOGEE project, APOGEE-2, includes  an expansion of the survey to the Southern Hemisphere called APOGEE-2S, which observes the sky simultaneously with APOGEE-2N, using a cloned spectrograph.
This enabled APOGEE to truly map the Milky Way in a panoramic manner, while putting special attention to the dust-hidden inner regions of the Milky Way, which is crucial to make comprehensive chemodynamical analysis of the Milky Way.

\citetalias{zasowski2017}, presented the 
targeting selection strategy used for APOGEE-2 (both North and 
South counterparts), which has been modified and updated as the survey has progressed.
Throughout this paper we present the final targeting strategy of
APOGEE-2S, with special attention to the changes applied since
\citetalias{zasowski2017}.
First, we briefly described the main concepts involved in the process of target selection \added{for all components of APOGEE.}
\explain{ From this point forward the section was rewritten entirely}
Then, we presented the main motivations for changing our target selection with respect to the one presented in \citetalias{zasowski2017}, which involved internal evaluations to maximize our ability to reach our science goals, the use of data from new external surveys in our selection methods, and the alterations of our observing schedule. Probably the latter is the most important motivation for this paper since our original plan was designed for 1493 visits obtained throughout 320 observing nights. However, as explained in \autoref{sec:motivation}, our observing efficiency, and the observing schedule changed significantly over the course of the survey. The COVID-19 pandemic global spread in early 2020, implied the lost of 77 observing nights, which was compensated by the extension that was granted to APOGEE-2S, that included 88 nights granted for the end of year 2020. With all these changes our final observing strategy encompassed 2120 total planned visits,  obtained throughout 352 observing nights, which represent $42\%$ more than the original observing plan.

The other subsections of \autoref{sec:mainchanges} summarized the main changes applied to the constituent programs of the main survey. The bulge program was unintentionally filled with stars fainter than needed given our planned visits per field, but thanks to a modification in the plan and increment in our observing efficiency, we obtained a bulge coverage that was significantly deeper than our original bulge program. For the halo, thanks to the extra nights granted, we could add new fields meant to target both halo and stellar stream targets, and using the results obtained from APOGEE2-N we could improve our halo selection method and increase our efficiency to target distant stars. Other project that was significantly modified was the Magellanic Cloud program for which we practically doubled the number of fields in our observing plan thanks to the nights allocated for the end of year 2020. Other changes to our observing plan involve the inclusion of the Fornax dSph galaxy, and new globular and open clusters that allowed us to fine-tune our metallicity calibration.
We also present our definitive observing field plan map, and provide the final version of the list of targeting flags used  in this survey, for the user to identify the purpose for which all the different targets of APOGEE-2S were intended.

Finally, we present for the first time the list of contributed programs to APOGEE-2S, which are allocated either through the Chilean Time Allocation Committee (CNTAC) or by the Observatories of the Carnegie Institution of Science (CIS) TAC. These 19 programs are designed by the PI responsible and the APOGEE-2S team only perform checks of observability and pluggability. Each program is scheduled for specific nights, and the total nights assigned to contributed programs in APOGEE-2S was 104, which represents $\sim23\%$ of the total observing time of the survey. The contributed program section provides small descriptions of each project along with the contact information from the PI, whereas the overall coverage of contributed programs is shown in \autoref{fig:contributed_distribution}.
\begin{acknowledgements}
The APOGEE project thanks Jeff Munn (NOFS) for collecting Washington$+DDO51$ imaging for large areas of the sky. 
The authors also recognize contributions from prior members of the targeting team and the members of the science working groups in both APOGEE-1 and APOGEE-2. 
We graciously thank the staff at Las Campanas Observatory and the APOGEE-2S Observing team for both their dedication and their remarkable ability to exceed themselves without needing to. 
We also warmly recognize the APOGEE-S Operations team for their diligence, the Carnegie Observatories Telescope Allocation staff, the members of the CNTAC, and the SDSS Plate Shop for their enduring patience.
We thank the SDSS-IV Data Team for thinking about and then handling so many details and doing so without a complaint, albeit often with a stroopwafel. 
We thank the SDSS-IV Management Committee and all SDSS-IV leadership whose commitment to SDSS-IV is literally is the glue that keeps the project going.
Lastly, we warmly thank Jim Gunn and Jill Knapp for the remarkable manner in which they have and continue to impact the world around them.
%

F.A.S. acknowledges support from CONICYT Project AFB-170002.
Support for this work was provided by NASA through Hubble Fellowship grant \#51386.01 awarded to R.L.B. by the Space Telescope Science Institute, which is operated by the Association of  Universities for Research in Astronomy, Inc., for NASA, under contract NAS 5-26555.
K.R.C acknowledges support provided by the NSF through grant AST-1449476.
S.R.M. acknowledges support through NSF grants AST-1616636 and AST-1909497.
The research leading to these results has (partially) received funding from the European Research Council (ERC) under the European Union's Horizon 2020 research and innovation programme (grant agreement N$^\circ$670519: MAMSIE), from the KU~Leuven Research Council (grant C16/18/005: PARADISE), as well as from the BELgian federal Science Policy Office (BELSPO) through PRODEX grant PLATO (C.A. and A.T.).
J.B. and N.M. are supported by the Ministry for the Economy, Development and Tourism, Programa Iniciativa Cientica Milenio grant IC120009, awarded to the Millennium Institute of Astrophysics (MAS).
J.D. and P.M.F. acknowledge support for this research from the National Science Foundation AAG and REU programs (AST-1311835, AST-1715662, PHY-1358770, \& PHY-1659444).
J.G.F-T. is supported by FONDECYT No. 3180210 and Becas Iberoam\'erica Investigador 2019, Banco Santander Chile.
D.G. gratefully acknowledges support from the Chilean Centro de Excelencia en Astrof\'isica
y Tecnolog\'ias Afines (CATA) BASAL grant AFB-170002.
D.G. also acknowledges financial support from the Direcci\'on de Investigaci\'on y Desarrollo de la Universidad de La Serena through the Programa de Incentivo a la Investigaci\'on de Acad\'emicos (PIA-DIDULS).
S.H. is supported by an NSF Astronomy and Astrophysics Postdoctoral Fellowship under award AST-1801940.
D.M. acknowledges support by the BASAL Center for Astrophysics and Associated Technologies (CATA) through grant AFB 170002, and by Proyecto FONDECYT No. 1170121. 
R.R.M. acknowledges partial support from project BASAL AFB-$170002$ as well as FONDECYT project N$^{\circ}1170364$.
ARA acknowledges support from FONDECYT through grant 3180203.
A. Roman-Lopes acknowledges financial support provided in Chile by Comisi\'on Nacional de Investigaci\'on Cient\'ifica y Tecnol\'ogica (CONICYT) through the FONDECYT project 1170476 and by the QUIMAL project 130001. 
AMS gratefully acknowledges funding support through Fondecyt Regular (project code1180350) and funding support from Chilean Centro de Excelencia en Astrof\'isica y Tecnolog\'ias Afines (CATA) BASAL grant AFB-170002.
G.Z. acknowledges support from the NSF through grants AST-1203017 and AST-1911129.
%
Funding for the Sloan Digital Sky Survey IV has been provided by the Alfred P. Sloan Foundation, the U.S. Department of Energy Office of Science, and the Participating Institutions. SDSS-IV acknowledges support and resources from the Center for High-Performance Computing at the University of Utah. The SDSS web site is \url{ www.sdss.org}.

SDSS-IV is managed by the Astrophysical Research Consortium for the Participating Institutions of the SDSS Collaboration including the Brazilian Participation Group, the Carnegie Institution for Science, Carnegie Mellon University, the Chilean Participation Group, the French Participation Group, Harvard-Smithsonian Center for Astrophysics, Instituto de Astrof\'isica de Canarias, The Johns Hopkins University, Kavli Institute for the Physics and Mathematics of the Universe (IPMU) / University of Tokyo, Lawrence Berkeley National Laboratory, Leibniz Institut f\"ur Astrophysik Potsdam (AIP),  Max-Planck-Institut f\"ur Astronomie (MPIA Heidelberg), Max-Planck-Institut f\"ur Astrophysik (MPA Garching), Max-Planck-Institut f\"ur Extraterrestrische Physik (MPE), National Astronomical Observatories of China, New Mexico State University, New York University, University of Notre Dame, Observat\'ario Nacional / MCTI, The Ohio State University, Pennsylvania State University, Shanghai Astronomical Observatory, United Kingdom Participation Group, Universidad Nacional Aut\'onoma de M\'exico, University of Arizona, University of Colorado Boulder, University of Oxford, University of Portsmouth, University of Utah, University of Virginia, University of Washington, University of Wisconsin, Vanderbilt University, and Yale University.

This publication makes use of data products from the Two Micron All Sky Survey, which is a joint project of the University of Massachusetts and the Infrared Processing and Analysis Center/California Institute of Technology, funded by the National Aeronautics and Space Administration and the National Science Foundation.  This publication also makes use of data products from the Wide-field Infrared Survey Explorer, which is a joint project of the University of California, Los Angeles, and the Jet Propulsion Laboratory/California Institute of Technology, funded by the National Aeronautics and Space Administration.

This work is based, in part, on observations made with the \spitzer\ Space Telescope, which is operated by the Jet Propulsion Laboratory, California Institute of Technology under a contract with NASA.

This publication makes use of data products from the Wide-field Infrared Survey Explorer \citep{wright2010}, which is a joint project of the University of California, Los Angeles, and the Jet Propulsion Laboratory/California Institute of Technology, funded by the National Aeronautics and Space Administration, and NEOWISE, which is a project of the Jet Propulsion Laboratory/California Institute of Technology. 
WISE and NEOWISE are funded by the National Aeronautics and Space Administration 

This work has made use of data from the European Space Agency (ESA) mission \gaia\ (\url{https://www.cosmos.esa.int/gaia}), processed by the \gaia\ Data Processing and Analysis Consortium (DPAC, \url{https://www.cosmos.esa.int/web/gaia/dpac/consortium}). Funding for the DPAC has been provided by national institutions, in particular the institutions participating in the \gaia\ Multilateral Agreement.

This research has made use of NASA’s Astrophysics Data System.

Many of the acknowledgements were compiled using the Astronomy Acknowledgement Generator. 

This research has made use of the SIMBAD database, operated at CDS, Strasbourg, France. The original description of the SIMBAD service was published in \citet{simbad_2000}.

This research has made use of the \vizier\ catalogue access tool, CDS, Strasbourg, France (DOI: 10.26093/cds/vizier). The original description of the \vizier\ service was published \citet{vizier2000}.
\end{acknowledgements}

\facilities{Du Pont (APOGEE), Sloan (APOGEE), \spitzer, \emph{WISE}, 2MASS, SkyMapper, \gaia}

\software{ 
    \replaced{Astropy, NumPy, Interactive Data Language (IDL), TopCat}{Astropy \citep{Astropy_2013,Astropy_2018}, 
    NumPy \citep{vanderWalt_2011,Harris_2020numpy},
    TopCat \citep{topcat}}
    }

\bibliographystyle{aasjournal}
 \newcommand{\noop}[1]{}

\begin{appendix} 
\section{GLOSSARY} \label{glossary}   
This Glossary contains SDSS- and APOGEE-specific terminology appearing in this paper and throughout the data documentation. 

\begin{description} \itemsep -2pt
\item[1-Meter Target] Target observed with the NMSU 1-m telescope (\texttt{TELESCOPE} tag of `apo1m'), which has a single fiber connection to the APOGEE-2N instrument. The NMSU 1-m telescope is described in \citet{holtzman_2010} with the reduction specific to its connection to the APOGEE-N instrument given in \citet{holtzman2015}.
\item[Ancillary Target] Target observed as part of an approved Ancillary Science Program. Ancillary science Programs from APOGEE-1 are described in \citetalias{zasowski2013} and from APOGEE-2 in \beatonetal. 
\item[APO] Apache Point Observatory; site of the Sloan Foundation 2.5-m telescope \citep{gunn2006} on which the APOGEE-N spectrograph operates. 
\item[ASPCAP] The APOGEE Stellar Parameters and Chemical Abundances Pipeline; the analysis software that calculates basic stellar parameters (T$_{\rm eff}$, $\log{g}$, [Fe/H], [$\alpha$/Fe], [C/Fe], [N/Fe]) and elemental abundances \citep{holtzman2015,garciaperez2016}.
\item[BTX] The Bright Time Extension, an APOGEE-2N program executed in the last 1.5 years of the APOGEE-2 Survey.
\item[CIS] The Carnegie Institution for Science or CIS is an SDSS-IV partner and operates the Las Campanas Observatory in Chile. 
\item[Cohort] Set of targets in the same field that are observed together on all of their visits.  A given plate may have multiple cohorts on it.
\item[Contributed Program] Term for programs allocated to Principal Investigators by the CIS or CNTAC but whose data are contributed to the APOGEE-2 survey. These data appear in SDSS data releases, but their targeting was performed by the PI. These programs are described in this work.
\item[CNTAC] The Chilean National Telescope Allocation Committee, which allocates observing resources to the Chilean community. 
\item[Design] Set of targets drilled together on a plate, consisting of up to one each of short, medium, and long cohorts.  A design is identified by an integer Design ID. Changing a single target on a design results in a new design.
\item[Design ID] Unique integer assigned to each design.
\item[Drill Angle] Hour angle (distance from the meridian) at which a plate is drilled to be observed.  This places the fiber holes in a way that accounts for differential refraction across the FOV.
\item[External Program] General term for programs and targets observed during the APOGEE-2S time allocated by the Carnegie Observatories (OCIS) or the Chilean Time Allocation Committee (CNTAC). These targets will not be included in the SDSS dataset.
\item[Fiber Collision] A situation in which two targets, separated by less than the protective ferrule around the fibers, are included in the same design.  The higher-priority target is drilled on the plate(s); the lower-priority target is removed.
\item[Fiber ID] Integer (1--300) corresponding to the rank-ordered spectrum on the detector. Fiber IDs can vary from visit to visit for a given star.
\item[Field] Location on the sky, defined by central coordinates and a plate radius.
\item[GAP] The \ktwo\ Galactic Astrophysics Program, which is described in \citet{Stello_2017}.
\item[LCO] Las Campanas Observatory, site of the Ir\'en\'ee~du~Pont $2.5$-${\rm m}$ telescope \citep{bowen1973} on which the APOGEE-S spectrograph operates. 
\item[Location ID] Unique integer assigned to each field on the sky.
\item[Main Red Star Sample] The sample drawn from a simple selection function defined by magnitude and color that comprises the bulk of the APOGEE program. This program is explained in \citetalias{zasowski2013} and \citetalias{zasowski2017}.
%
%
\item[MaNGA] Mapping Nearby Galaxies at Apache Point Observatory;  A SDSS-IV program described in \citet{bundy_2015}. 
\item[MaStar] The MaNGA Stellar Program; a program within the MaNGA Survey with the objective of constructing a high-fidelity stellar library. An overview of the project and its first data release is described in \citet{yan_2019}.
\item[POI] Photometric Object of Interest; an umbrella term for stars targeted due to their \kepler, \ktwo, or \tess\ light curves.
\item[Plate] Piece of aluminum with a design drilled into it.  Note that while ``plate'' is often used interchangeably with ``design'', multiple plates may exist for the same design -- e.g., plates with a common design but drilled for different hour angles.
\item[Plate ID] Unique integer assigned to each plate.
\item[RJCE] The Rayleigh-Jeans Color Excess method, a technique used to estimate the line-of-sight reddening to a star \citep{majewski2011}.  APOGEE-2 uses this method to estimate intrinsic colors for many potential targets \citepalias[for details see][]{zasowski2013,zasowski2017}.
\item[Sky Targets] Empty regions of sky on which a fiber is placed to collect a spectrum 
used to remove the atmospheric airglow lines and sky background from the target spectra observed simultaneously with the same plate.
\item[Special Targets] General term for targets selected with criteria other than the color and magnitude criteria of the main red giant sample.  For example, special targets include ancillary science program targets and calibration cluster members.
\item[Targeting Flag and Bits] A targeting ``flag'' refers to one of the three long integers assigned to every target in a design, each made up of 31 ``bits'' that correspond to particular selection or assignment criteria.  APOGEE-2's flags are named \texttt{APOGEE2\_TARGET1}, \texttt{APOGEE2\_TARGET2}, \texttt{APOGEE2\_TARGET3}, and \texttt{APOGEE2\_TARGET4}; see Table~\ref{tab:targeting_bits} for a list of the bits as of this publication. 
\item[Telluric Standards] Hot blue stars observed on a plate to derive corrections for the telluric absorption lines.
\item[Visit] The base unit of observation, equivalent to approximately one hour of on-sky integration (but this can vary, see \autoref{sec:concepts}) and comprising a single epoch.  Repeated visits are used to both build up signal and provide a measure of spectral and RV stability.
\item[Washington\textit{+DDO51}] Also ``W+D photometry''; adopted abbreviation for the combination of Washington $M$ and $T_2$ photometry \citep{Canterna_1976} with $DDO51$ photometry \citep{McClure_1973}, used in the photometric classification of dwarf/giant stars \citep{majewski2000}.
\end{description}
\end{appendix} 

\end{document}